\documentclass[11pt]{article}

\usepackage[T1]{fontenc}
\usepackage[cp1251]{inputenc}
\usepackage{textcomp}
\usepackage[centertags]{amsmath}
\usepackage[mediummath]{nccmath}
\usepackage{amsfonts}
\usepackage{amssymb}
\usepackage{mathtools}
\usepackage[pdftex]{hyperref}
\usepackage{graphicx}
\usepackage{graphbox}
\usepackage[numbers,sort&compress]{natbib}

\usepackage{paperinitial}

\paperinitialization{15mm}{15mm}{15mm}{15mm}{2pt}{10pt}

\DeclareMathOperator{\sgn}{sgn}

\newcommand{\vf}{\varphi}
\newcommand{\vk}{\varkappa}

\newcommand{\al}{\alpha}
\newcommand{\be}{\beta}
\newcommand{\ga}{\gamma}

\newcommand{\de}{\delta}
\newcommand{\De}{\Delta}

\newcommand{\la}{\lambda}

\newcommand{\spx}{\mathbf{x}}

\newcommand{\spk}{\mathbf{k}}

\newcommand{\N}{\mathbb{N}}
\newcommand{\Z}{\mathbb{Z}}

\begin{document}
	\allowdisplaybreaks[4]
	\frenchspacing

\title{{\Large\textbf{Radiation of twisted photons in elliptical multifrequency undulators}}}

\date{}

\author{%
O.V. Bogdanov${}^{1)}$\thanks{E-mail: \texttt{bov@tpu.ru}},\;
S.V. Bragin${}^{1)}$\thanks{E-mail: \texttt{svb38@tpu.ru}},\;
P.O. Kazinski${}^{2)}$\thanks{E-mail: \texttt{kpo@phys.tsu.ru}},\;
and
V.A. Ryakin${}^{1,2)}$\thanks{E-mail: \texttt{vlad.r.a.phys@yandex.ru}}\\[0.5em]
{\normalsize ${}^{1)}$Mathematics and Mathematical Physics Division,}\\
{\normalsize Tomsk Polytechnic University, Tomsk 634050, Russia}\\[0.5em]
{\normalsize ${}^{2)}$ Physics Faculty, Tomsk State University, Tomsk 634050, Russia}
}

\maketitle

\begin{abstract}

The theory of radiation of twisted photons in elliptical multifrequency undulators is developed. It is shown that helical multifrequency undulators can be employed as a bright and versatile source of photons in the states that are superpositions of the modes with definite projection of total angular momentum (TAM), amplitude, and relative phase. All these parameters of the state are readily controlled by the undulator design. The explicit expression for the amplitude of radiation of a twisted photon from a charged particle in the multifrequency undulator is derived. The energy spectrum of radiation and the selection rules for the TAM projection of radiated photons are described. The symmetry properties of the spectrum with respect to the TAM projection are established. The interpretation to the energy spectrum and to the selection rules is given in terms of virtual photons mediating between the charged particle and the undulator. The results obtained are also applicable to radiation of twisted photons produced by ultrarelativistic charged particles moving in plane multifrequency electromagnetic waves.

\end{abstract}

\section{Introduction}

The states of photons that are coherent superpositions of modes with definite projection of total angular momentum (TAM) find their application in studying interference effects in various processes of quantum electrodynamics \cite{Stock2015,Sherwin2017,Sherwin2020,Ivanov2022}. The photons in such states excite coherently the different rotational modes of quantum systems such as elementary particles, atoms, excitons, and nuclei \cite{Ivanov2022,Roadmap16,KnyzSerb,Afanas13,Mukherjee2018,Lange2022,Peshkov2023,KKR_Rydberg2025,QRTKrmp22,KazRyak23,KazSok2024,Lu2023} leading to interference effects that are absent in exciting these systems by photons in the states with definite projection of TAM, including plane wave photons. In particular, in scattering of a twisted photon, i.e., the photon with definite projection of TAM on its propagation axis \cite{Roadmap16,KnyzSerb,ZWYB20}, on particles described by plane wave states, the inclusive probability to record some outgoing particles in the states with definite momenta is an incoherent sum of scattering probabilities for particles prepared and detected in the plane wave states \cite{Ivanov2022}. Information about the nontrivial phase of the twisted photon is lost. Apart from this application, the beams of photons prepared in such superposed states are used in optical tweezers for manipulation of polarizable microparticles, their rotation, and confining at assigned points \cite{Franke-Arnold2007}. These states of photons are also employed to increase resolution in optical metrology \cite{Cheng2025}. In the present paper, we show that multifrequency undulators provide a bright source of photons in such states. The photon frequency, the relative phases of twisted modes in the quantum state, and their amplitudes are controlled by the undulator parameters.

The multifrequency undulators are a classical subject of a theory of radiation by relativistic particles \cite{Iracane1991,Ciocci1993,Bazylev1993,BaKaStrbook,Bord.1,Zhukovsky2021}. Nevertheless, the theory of radiation of twisted photons in such undulators has apparently not been developed yet. The present paper aims to fill this gap. Even in studying radiation of plane wave photons by such undulators, the theory was constructed only for several particular cases with the number of frequencies not greater than four \cite{Zhukovsky2021,Tripathi2011,Zhukovsky2020,ZhukovskyFedorov2021,Gajbhiye2023}. The two-frequency undulators and the free-electron lasers based on them are the most completely investigated \cite{Iracane1991,Ciocci1993,Bazylev1993,Dattoli2006,Mishra2009,Mirian2014,Datolli2014,Zhukovsky2016,Zhukovsky2017nano,Zhukovsky2017, Zhukovsky2017epl,Mishra2020}. In particular, it is shown in these papers that, varying the parameters of multifrequency undulator, one can enhance or suppress the radiation intensity at specified harmonics and change the polarization properties of radiation. In the present paper, we develop the theory of radiation of twisted photons by a charged particle moving in an undulator composed of $M$ arbitrary elliptical undulators with different frequencies, i.e., the magnetic field in such an undulator is a linear superposition of the magnetic fields of $M$ elliptic one-frequency undulators, and obtain the explicit expression for the one-particle amplitude of radiation of a twisted photon. A special attention is paid to the case when the ratios of undulator frequencies are rational numbers. In that case, the energy spectrum of radiated photons and the selection rules for the TAM projection are completely described with the aid of the B\'{e}zout coefficients. One of the consequences of this analysis is that the radiation frequency of the lowest harmonic is always less than the radiation frequencies of one-frequency undulators constituting the multifrequency undulators unless all the frequencies of the multifrequency undulator are multiples of one of the frequencies of the one-frequency undulators comprising the multifrequency one. In the present paper, we investigate only the radiation from a single charged particle. A generalization of the theory to beams of charged particles is implemented along the lines of the papers \cite{Bogdanov2019,Bogdanov2020}.

In the ultrarelativistic case, the properties of radiation from charged particles moving in an undulator are similar to the properties of radiation from charged particles propagating in a plane electromagnetic wave. In this connection, we mention the papers \cite{Taira2018,Jiang2025} where the investigation of radiation of twisted photons by charged particles moving in a plane two-color electromagnetic wave was carried out. In particular, the general selection rules for the TAM projection of radiated twisted photons were established in the case when the electromagnetic wave is a superposition of two circularly polarized waves with different frequencies. However, even in this two-frequency case a detailed analysis of the radiation spectrum was not performed when the ratio of frequencies is a rational number. For such a case, the selection rule for the TAM projection at a given radiation harmonic, i.e., at a fixed energy of a radiated photon, was not obtained in these papers. Nor was the explicit expression for the state of radiated photons derived. We will show below that it is a coherent superposition of twisted modes with different projections of TAM. Among the other means to generate photons in the states being a superposition of twisted modes, we mention the ring-shaped phase plates with different topological charges \cite{Vasilyeu2009}, the specially designed metasurfaces \cite{Zheng2021,Shi2025}, and the diffraction gratings \cite{Lee2019}. A significant drawback of these approaches is the low brightness of these sources as compared to undulator radiation and a narrow frequency band where the photons in the desired state are generated. Notice also the paper \cite{Liu2024} where the tandem of undulators was used to generate coherent superpositions of Laguerre-Gaussian modes with opposite projections of the orbital angular momentum.

The paper is organized as follows. In Sec. \ref{General_Formulas}, we provide the expressions for the trajectory of a charged particle in multifrequency undulator and the general formulas for the average number of radiated twisted photons. In Sec. \ref{Probab_Rad_Tw_Phot}, we derive the explicit expression for the amplitude of radiation of a twisted photon in multifrequency undulator. There we also analyze the energy spectrum of radiated photons, deduce the selection rules for TAM projection, and prove some symmetry properties of the average number of radiated photons. Section \ref{Two-frequency_Undulator} is devoted to two-frequency undulators. We particularize the general formulas to this case and obtain the explicit expression for the state of radiated photons. Moreover, we find the selection rule for the TAM projection of photons radiated by a helical $M$-frequency undulator at the $n$-th harmonic and obtain a more general selection rule for the TAM projection of photons radiated by an arbitrary $M$-frequency undulator. In Conclusion, we summarize the results.

We use the notation and agreements adopted in the papers \cite{BKL2,BKL4,BKL6,KazRyakElUnd}. In particular, we use the system of units such that $\hbar=c=1$ and $e^2=4\pi\al$, where $\al$ is the fine structure constant. We also assume that $x\equiv x_1$, $y\equiv x_2$, and $z\equiv x_3$.

\section{General formulas}\label{General_Formulas}

In this section, we present the general formulas describing the trajectories of charged particles in an $M$-frequency undulator and the average number of twisted photons emitted by such particles. Consider a system that is a composition of $M$ elliptical undulators with section lengths $ l_i $. Let us assume that an ultrarelativistic charged particle with charge $e$ moves on average along the $z$ axis in the positive direction. In the vicinity of the $z$ axis, there is a stationary magnetic field of the form (see, e.g., \cite{BaKaStrbook,Bord.1,Zhukovsky2021})
\begin{equation}\label{magnetic_field}
    H_x=\sum_{i=1}^M H_x^i\sin \tilde{\vf}_i,\qquad H_y=\sum_{i=1}^M H_y^i\cos \tilde{\vf}_i,\qquad H_z=0,
\end{equation}
where $\tilde\vf_{i}=\pm2\pi z/l_i +\chi_{i}$, and $H_x^i$, $H_y^i$, and $\chi_i$ are some constants. Since the particle moves in a constant magnetic field, its Lorentz factor, $\ga$, is an integral of motion. Integrating the Lorentz equations in the ultrarelativistic approximation, $\ga\gg1$, we obtain
\begin{equation}\label{velocities}
\begin{gathered}
    \dot{x}=-\sum_{i=1}^M a_i\omega_i \sin\vf_i,\qquad \dot{y}=\sum_{i=1}^M b_i\omega_i \cos\vf_i,\\ \dot{z}=\sqrt{1-\ga^{-2}-\dot{x}^2-\dot{y}^2}\approx \be_3 +\sum_{i=1}^M\frac{\omega_i^2}{4} (a_i^2-b_i^2)\cos(2\vf_i)-\\
    -\sideset{}{'}\sum_{i,j=1}^M \frac{\omega_i\omega_j}{4}\big[(a_ia_i+b_ib_j)\cos(\vf_i-\vf_j) -(a_ia_j -b_ib_j)\cos(\vf_i+\vf_j)\big],
\end{gathered}
\end{equation}
where the dot denotes the derivative with respect to the laboratory time $t$, the prime near the sum sign means that the terms with $i=j$ are omitted, $\be_3$ is the average particle velocity along the $z$ axis, which has the form
\begin{equation}\label{dip_parameter}
    \be_3=1-\frac{1+K^2}{2\ga^2},\qquad K^2=\sum_{i=1}^M K_i^2,\qquad K_i^2=\frac{\omega_i^2\ga^2}{2}(a_i^2+b_i^2),
\end{equation}
and $\vf_{i}=\omega_{i} t +\chi_{i}$, where $\omega_{i}=\pm2\pi\be_3/l_{i}$. The sign of $\omega_i$ determines the chirality of the $i$-th elliptical undulator. For definiteness, we assume that $|\omega_1|$ is minimal among all $|\omega_i|$. The following notation has also been introduced
\begin{equation}\label{a_b}
    a_i=\frac{eH_y^i}{m_e\ga\omega_i^2},\qquad b_i=-\frac{eH_x^i}{m_e\ga\omega_i^2},
\end{equation}
where $m_e$ is the mass of the charged particle (the electron). On integrating equations \eqref{velocities}, we obtain
\begin{equation}\label{trajectory}
	x= x_0+ \sum_{i=1}^{M}a_{i} \cos\vf_i,\qquad y=y_0+ \sum_{i=1}^{M} b_i\sin\vf_i, \qquad z= z_0+\be_3 t+\sum_{i,j=1}^M
    \big[c_{ij}\sin(\vf_i+\vf_j)+d_{ij}\sin(\vf_i-\vf_j)\big],
\end{equation}
where $x_0$, $y_0$ and $z_0$ are some constants. The amplitudes of longitudinal oscillations of the particle in the undulator are written as
\begin{equation}\label{c_d}
	c_{ij}=c_{ji}=\frac{\omega_i\omega_j}{4}\frac{a_ia_j-b_ib_j}{\omega_i+\omega_j},\qquad
    d_{ij}=-d_{ji}=-\frac{\omega_i\omega_j}{4}\frac{a_ia_j+b_ib_j}{\omega_i-\omega_j}.
\end{equation}
By definition, we set $d_{ii}=0$. Expression \eqref{trajectory} is valid under the assumption that $\ga\gg1$ and $K_{i}/\ga\ll1$. We suppose that these estimates are valid. The dipole case, when all $K_i\ll1$, has been studied in detail in \cite{BKL2,BKL4}. Therefore, in this paper, we are interested in the case when there is at least one $K_i\gtrsim1$ among $K_i$.

A charged particle moves along the trajectory \eqref{trajectory} for $t\in[-TN/2,TN/2]$, where $TN:=L/\beta_3$, $L$ is the length of the undulator, and $N\gg1$ is the number of its sections for the minimum frequency $|\omega_1|$. When $t$ does not belong to the interval specified, we will assume that the particle moves parallel to the $z$ axis with the velocity $\be_\parallel=\sqrt{1-1/\ga^2}$. We are interested in radiation from the part of the trajectory $t\in[-TN/2,TN/2]$. This radiation dominates when $N\gg1$ for the photon energies corresponding to harmonics of the undulator radiation.

The average number of twisted photons radiated by a classical point charge is given by \cite{BKL2}
\begin{multline}\label{probabil}
	dP(s,m,k_3,k_\perp)=e^2\bigg|\int d\tau
    e^{-i[k_0x^0(\tau)-k_3x_3(\tau)]}\Big\{\frac12\big[\dot{x}_+(\tau)a_-(s,m,k_3,k_\perp;\spx(\tau))+\\
	+\dot{x}_-(\tau)a_+(s,m,k_3,k_\perp;\spx(\tau)) \big]
	+\dot{x}_3(\tau)a_3(m,k_\perp;\spx(\tau))\Big\} \bigg|^2 n_\perp^3\frac{dk_3dk_\perp}{16\pi^2},
\end{multline}
where $s$ is the helicity of the twisted photon, $m$ is the TAM projection of the twisted photon onto the $z$ axis, $k_\perp$ and $k_3$ are the corresponding projections of the momentum of the twisted photon, $k_0=\sqrt{k_3^2+k_\perp^2} $ is the photon energy, $n_\perp:=k_\perp/k_0$ (see for details \cite{JenSerprl,JenSerepj,BKL2,BKL4,BKL6}). We have also introduced the notation for the components of particle trajectory
\begin{equation}
	x_\pm:=x\pm iy.
\end{equation}
The mode functions of twisted photons $a_\pm$, $a_3$ can be cast into the form
\begin{equation}\label{mode_func_an}
\begin{split}
   a_3&\equiv a_3(m, k_\perp;\spx)=J_{m} (k_{\bot} |x_{+}|) e^{i m \arg( x_{+})}=:j_m(k_\perp x_+,k_\perp x_-),\\
    a_{\pm}&\equiv a_{\pm} (s,m,k_3, k_\perp; \spx)=i\frac{s \mp \tilde{n}_3}{n_\perp}  j_{m\pm1}(k_\perp x_+,k_\perp x_-),
\end{split}
\end{equation}
where $\tilde{n}_3:=k_3/k_0$. The parameter $\tau$ in \eqref{probabil} is an arbitrary parameter on the particle worldline. In our case, it is convenient to choose it equal to the laboratory time $\tau=t\equiv x^0(t)$.

\section{Probability of radiation of twisted photons}\label{Probab_Rad_Tw_Phot}

To obtain the amplitude of radiation and the average number of emitted twisted photons, it is necessary to calculate the integrals entering into \eqref{probabil}. Let us introduce the notation
\begin{equation}\label{I_integrals}
\begin{split}
	I_3&:=\int_{-TN/2}^{TN/2}dt e^{-ik_0[t -\tilde{n}_3(z_0+\be_3 t+
    \sum_{i,j=1}^{M}[c_{ij}\sin(\vf_i+\vf_j) +d_{ij}\sin(\vf_i-\vf_j)])]}\dot{x}_3a_3(m,k_\perp;\spx(t)),\\
	I_\pm&:=\int_{-TN/2}^{TN/2}dt e^{-ik_0[t -\tilde{n}_3(z_0+\be_3 t+\sum_{i,j=1}^{M}[c_{ij}\sin(\vf_i+\vf_j)
    +d_{ij}\sin(\vf_i-\vf_j)])]}
    \dot{x}_\pm a_\mp(s,m,k_3,k_\perp;\spx(t)).
\end{split}
\end{equation}
For $N\gg1$, these integrals give the main contribution to \eqref{probabil}, i.e., the contribution of edge radiation to \eqref{probabil} can be neglected. The average number of radiated twisted photons becomes
\begin{equation}\label{probab_I}
	dP(s,m,k_3,k_\perp)\approx e^2|\mathcal{A} |^2 n_\perp^3\frac{dk_3dk_\perp}{16\pi^2},
\end{equation}
where
\begin{equation}
    \mathcal{A}:=I_3+(I_++I_-)/2
\end{equation}
is proportional to the one-particle amplitude of radiation of a twisted photon by a classical current. Recall that the intensity of radiation of twisted photons with given quantum numbers equals $k_0 dP(s,m,k_3,k_\perp)$. The components of the particle trajectory \eqref{trajectory} have the form
\begin{equation}
	x_\pm=x_\pm^0+\sum_{i=1}^M (R_i e^{\pm i\vf_i} + D_i e^{\mp i\vf_i}),
\end{equation}
where
\begin{equation}\label{R_D_x0}
	R_i := \frac{a_i + b_i}{2}, \qquad D_i := \frac{a_i - b_i}{2}, \qquad x_\pm^0 := x_0 \pm i y_0.
\end{equation}
The velocities are written as
\begin{equation}\label{dotxpm}
\begin{gathered}
	\dot{x}_\pm = \pm i \sum_{i=1}^M \omega_{i}(R_ie^{\pm i\vf_i} - D_i e^{\mp i\vf_i}),\\
    \dot{x}_3=\be_3+\sum_{i,j=1}^M
    \big[(\omega_i+\omega_j)c_{ij}\cos(\vf_i+\vf_j)+(\omega_i-\omega_j)d_{ij}\cos(\vf_i-\vf_j)\big].
\end{gathered}
\end{equation}
As can be seen from formula \eqref{velocities}, the terms in the sum in the expression for $\dot{x}_3$ give a small contribution. We will neglect these terms.

Consider the integral $I_3$. It is convenient to evaluate it using the addition theorem for the Bessel functions (see (A6) in \cite{BKL2}) in the form
\begin{equation}\label{add_thm}
	j_\nu(x_++y_+,x_-+y_-)=\sum_{n=-\infty}^\infty j_{\nu-n}(x_+,x_-)j_n(y_+,y_-),
\end{equation}
and the property
\begin{equation}\label{j_prop}
	j_m(ap,q/a)=a^m j_m(p,q),\qquad m\in \mathbb{Z}.
\end{equation}
The definition of the functions $j_\nu(p,q)$ and their properties are given in Appendix A of \cite{BKL2}. By repeatedly applying these relations and redefining the summation indices accordingly, we obtain
\begin{multline}\label{h_3}
	j_{m}(k_\perp x_+,k_\perp x_-)\prod_{i,j=1}^{M}e^{ik_3[ c_{ij} \sin(\vf_i+\vf_j)
    +d_{ij}\sin(\vf_i-\vf_j)]}= \sum_{\{n_i\},\{r_i\},\{p_{ij}\},\{q_{ij}\}=-\infty}^\infty
     j_{m-\tiny{\sum_i} (n_i + 2 r_i) -2 \tiny{\sum_{ij}} p_{ij}}^0   \times\\
	\times \prod_{k,l=1}^M\Big[J_{q_{kl}}(\De_{kl}) J_{p_{kl}}(\vk_{kl}) \Big] \prod_{i=1}^{M}
    e^{in_i\vf_i}J_{n_i+r_i-\tilde{q}_i-\tilde{p}_i}(\rho_i)
    J_{r_i}(\de_i),
\end{multline}
where the exponents on the left-hand side have been expanded in the standard way into Fourier series over $\vf_i+\vf_j$ and $\vf_i-\vf_j$ with coefficients in the form of Bessel functions, the summation sign on the right of the equality sign in \eqref{h_3} implies summation over all the sets of the respective indices, and
\begin{equation}\label{tilde_q_tilde_p}
    \tilde{q}_i:=\sum_{j=1}^M(q_{ij}-q_{ji}),\qquad \tilde{p}_i:=\sum_{j=1}^M(p_{ij}+p_{ji}).
\end{equation}
Also, for brevity, the following notation has been introduced
\begin{equation}\label{j0_rho_delta_vk_Delta}
	j^0_m:=j_m(k_\perp x_+^0,k_\perp x_-^0),\qquad \rho_i:=k_\perp R_i,\qquad \de_i:=k_\perp D_i,\qquad\vk_{ij}=k_3c_{ij}, \qquad
    \De_{ij}=k_3d_{ij}.
\end{equation}
It is clear that
\begin{equation}
    \sum_{i=1}^M\tilde{q}_i=0,\qquad\sum_{i=1}^M\tilde{p}_i=2\sum_{i,j=1}^M p_{ij}.
\end{equation}
Since $\De_{ii}=0$, then $J_{q_{ii}}(\De_{ii})=\de_{q_{ii},0}$, the sums over the indices $q_{ii}$ are absent, and  it is assumed everywhere that $q_{ii}=0$. The dependence of expression \eqref{h_3} on $t$ is contained in the exponents on the right-hand side of \eqref{h_3}. As a result, the integral over $t$ is readily performed
\begin{equation}\label{I_3}
\begin{split}
	I_3=&\,2\pi\be_3 \sum_{\{n_i\},\{r_i\},\{p_{ij}\},\{q_{ij}\}=-\infty}^\infty \de_N\big(k_0(1-\tilde{n}_3\be_3)
    -\small{\sum_i} n_i\omega_i \big)e^{ik_3z_0} j_{m-\tiny{\sum_i} (n_i+2r_i) + 2\tiny{\sum_{ij}} p_{ij}}^0 \times\\
	&\times \prod_{k,l=1}^M\Big[J_{q_{kl}}(\De_{kl}) J_{p_{kl}}(\vk_{kl}) \Big] \prod_{i=1}^{M} e^{in_i\chi_i}
    J_{n_i+r_i-\tilde{q}_i-\tilde{p}_i}(\rho_i)
    J_{r_i}(\de_i) J_{p_i}(\vk_i),
\end{split}
\end{equation}
where
\begin{equation}\label{delta-shaped_seq}
	\de_N(x):=\frac{\sin(TNx/2)}{\pi x}.
\end{equation}
The representation \eqref{h_3} of the integrand of $I_3$ is the key point of the derivation of the radiation amplitude of twisted photons. The same technique can be employed for the derivation of the radiation amplitude of plane wave photons. In this case, the expansion analogous to \eqref{h_3} looks as
\begin{multline}\label{plane_wave_expans}
	e^{-ik_\mu x^\mu(t)}= e^{i\spk\spx_0-ik_0t(1-\tilde{n}_3\be_3)} \sum_{\{n_i\},\{r_i\},\{p_{ij}\},\{q_{ij}\}=-\infty}^\infty
   	\prod_{k,l=1}^M\Big[J_{q_{kl}}(\De_{kl}) J_{p_{kl}}(\vk_{kl}) \Big]\times\\
   \times \prod_{i=1}^{M}
    e^{in_i\vf_i -i(n_i+2r_i-\tilde{p}_i)(\psi-\pi/2)}J_{n_i+r_i-\tilde{q}_i-\tilde{p}_i}(\rho_i)
    J_{r_i}(\de_i),
\end{multline}
where $\psi=\arg(k_1+ik_2)$. We postpone the investigation of radiation of plane wave photons in $M$-frequency undulators for a separate study.

The integrals $I_\pm$ are evaluated in a way similar to $I_3$. For these integrals, we have
\begin{equation}\label{I_pm}
\begin{split}
	I_\pm =& - 2 \pi\frac{\tilde{n}_3\pm s}{n_\perp} \sum_{\{n_i\},\{r_i\},\{p_{ij}\},\{q_{ij}\}=-\infty}^\infty
    \de_N\big(k_0(1-\tilde{n}_3\be_3)-\small{\sum_i} n_i\omega_i \big)e^{ik_3z_0} j_{m-\tiny{\sum_i} (n_i+2r_i) + 2\tiny{\sum_{ij}} p_{ij}}^0 \times\\
	&\times e^{i\tiny{\sum_i}n_i\chi_i} \sum_{j=1}^M \omega_j \big[R_j J_{n_j+r_j -\tilde{q}_j-\tilde{p}_j\mp 1}(\rho_j)
    J_{r_j}(\de_j) - D_j J_{n_j+r_j-\tilde{q}_j-\tilde{p}_j}(\rho_j) J_{r_j \mp 1}(\de_j)\big]\times\\
    &\times\prod_{k,l=1}^M\Big[J_{q_{kl}}(\De_{kl}) J_{p_{kl}}(\vk_{kl}) \Big] \prod_{i\neq j}^{M}  J_{n_i+r_i-\tilde{q}_i-\tilde{p}_i}(\rho_i) J_{r_i}(\de_i)  .
\end{split}
\end{equation}
Thus the one-particle amplitude of radiation of a twisted photon is proportional to
\begin{multline}\label{totAmpl}
	\mathcal{A}= 2\pi \sum_{\{n_i\},\{r_i\},\{p_{ij}\},\{q_{ij}\}=-\infty}^\infty
    \de_N\big(k_0(1-\tilde{n}_3\be_3)-\small{\sum_i} n_i\omega_i \big)e^{ik_3z_0} j_{m-\tiny{\sum_i} (n_i+2r_i) + 2\tiny{\sum_{ij}} p_{ij}}^0 \times\\
	\times  e^{i\tiny{\sum_i}n_i\chi_i} \prod_{k,l=1}^M\Big[J_{q_{kl}}(\De_{kl}) J_{p_{kl}}(\vk_{kl}) \Big]
    \Big\{\Big(\be_3 - \sum_{j=1}^{M} \frac{\tilde{n}_3
    \omega_j(n_j-\tilde{q}_j-\tilde{p}_j)}{k_\perp n_\perp} \Big) \prod_{i=1}^{M}  J_{n_i+r_i-\tilde{q}_i-\tilde{p}_i}(\rho_i) J_{r_i}(\de_i)-\\
	 -\sum_{j=1}^M \frac{s\omega_j}{k_\perp n_\perp}
    \big[\rho_j J'_{n_j+r_j-\tilde{q}_j-\tilde{p}_j}(\rho_j) J_{r_j}(\de_j) - \de_j J_{n_j+r_j-\tilde{q}_j-\tilde{p}_j}(\rho_j)
    J'_{r_j}(\de_j)\big] \prod_{i\neq j}^{M}  J_{n_i+r_i-\tilde{q}_i-\tilde{p}_i}(\rho_i) J_{r_i}(\de_i) \Big\}.
\end{multline}
For large $N$ this expression has sharp maxima (the radiation harmonics) at
\begin{equation}\label{enSpect}
	k_0=\frac{n_1\omega_1+ \cdots + n_M\omega_M }{1-\tilde{n}_3\be_3}=:n_1\tilde{\omega}_1 +\cdots+ n_M \tilde{\omega}_M,
\end{equation}
corresponding to a set of integers $n_i$ such that the expression on the right-hand side is greater than zero.

Let us analyze the properties of the radiation energy spectrum. We introduce the coefficients $ \eta_i = \omega_i/\omega_1 $, $i=\overline{1,M}$. Then \eqref{enSpect} can be rewritten as
\begin{equation}
	k_0=\frac{\omega_1 (n_1+ \eta_2 n_2 +\cdots + \eta_M n_M) }{1-\tilde{n}_3\be_3}.
\end{equation}
Let the coefficients $\eta_i$ be rational numbers, $ \eta_i = h_i/g_i$, where $h_i$ and $g_i$ are relatively prime integers and $g_i\in\N$. Then the numbers $\tilde{g}_i:=\eta_i g$, where $  g:= lcm (g_i)\in\N$, are integers. Let us take the greatest common divisor $d=gcd(\tilde{g}_i)\in\N$ from the set of integers $\tilde{g}_i$. As a result, we obtain the following expression for the radiation energy spectrum
\begin{equation}\label{enSpect1}
	k_0=\frac{d\omega_1}{g}\frac{n_1 g/d + n_2 \tilde{g}_2 /d+\cdots + n_M \tilde{g}_M/d }{1-\tilde{n}_3\be_3}=
    \frac{d\omega_1}{g} \frac{n_1 \la_1 + n_2 \la_2+\cdots + n_M \la_M }{1-\tilde{n}_3\be_3},\qquad \la_i:=\tilde{g}_i/d\in\Z.
\end{equation}
According to the fundamental theorem on the greatest common divisor (see, e.g., \cite{HasseBook}), the numerator of the second fraction in \eqref{enSpect1},
\begin{equation}\label{numerator}
    n_1 \la_1 + n_2 \la_2+\cdots + n_M \la_M,
\end{equation}
is an integer $n$, which takes all possible values in $\Z$ in changing the integers $n_i$. As long as $k_0>0$, only those values of \eqref{numerator} whose sign coincides with the sign of $\omega_1$ are physical. In this case, the radiation energy spectrum \eqref{enSpect1} has the same form as the radiation energy spectrum of a one-frequency undulator,
\begin{equation}
    k_0=\frac{\omega n}{1-\tilde{n}_3\be_3},
\end{equation}
with
\begin{equation}\label{effective_freq}
    \omega = d\omega_1 /g.
\end{equation}
It is clear that $\sgn n=\sgn\omega$. The number $n$ enumerates the harmonics of radiation, i.e., it determines the energy spectrum, and therefore can be called the principal quantum number. For each value of the principal quantum number, i.e., for each value of $k_0$ from the radiation spectrum, there is an infinite number of sets of numbers $n_i$ corresponding to a given $n$. These sets of numbers differ by the solution $\de n_i$ of the homogeneous equation
\begin{equation}
    \de n_1 \la_1 + \de n_2 \la_2+\cdots + \de n_M \la_M=0.
\end{equation}
It is easy to see that the absolute value of the fundamental frequency of the undulator $|\omega|\leqslant|\omega_1|=\min|\omega_i|$, and equality is achieved only when all $\eta_i\in\Z$. If one of the numbers $\eta_i$ is irrational, then, formally, by going through all the possible values of $n_i$, the energy of the radiated photons \eqref{enSpect1} will take any value $k_0>0$ \cite{Ciocci1993}. Let us stress that these statements only determine the possible positions of peaks in the radiation energy spectrum. The magnitude of these peaks is specified by the modulus squared of expression \eqref{totAmpl}. In particular, for rational $\eta_i$, the distribution of radiation intensity across the energy spectrum \eqref{enSpect1} differs from the same distribution for a one-frequency undulator. These properties can be clearly seen in Figs. \ref{Plot_3helical237}, \ref{Plot_2helical32}, \ref{Plot_2helical32m}, \ref{Plot_2planar}.

Expression \eqref{totAmpl} for the radiation amplitude can be rewritten in a more compact form
\begin{equation}\label{totAmpl1}
\begin{split}
	\mathcal{A}=&\, 2\pi \sum_{\{n_i\},\{r_i\},\{p_{ij}\},\{q_{ij}\}=-\infty}^\infty
    \de_N\big(k_0(1-\tilde{n}_3\be_3)-\small{\sum_i} n_i\omega_i \big)e^{ik_3z_0}
    j_{m-\tiny{\sum_i} (n_i+2r_i) + 2\tiny{\sum_{ij}} p_{ij}}^0 \times\\
	&\times \prod_{i=1}^{M} \Big[ J_{n_i+r_i-\tilde{q}_i-\tilde{p}_i}(\rho_i) J_{r_i}(\de_i) e^{in_i\chi_i}\Big]
    \prod_{k,l=1}^M\Big[J_{q_{kl}}(\De_{kl}) J_{p_{kl}}(\vk_{kl}) \Big]\times\\
    &\times\Big\{ \be_3 -
    \sum_{j=1}^{M}\frac{\omega_j}{k_\perp n_\perp} \Big[ \tilde{n}_3
    (n_j-\tilde{q}_j-\tilde{p}_j) +s\frac{\partial}{\partial \ln k_\perp} \ln\frac{J_{n_j+r_j-\tilde{q}_j-\tilde{p}_j}(\rho_j)}{J_{r_j}(\de_j)}  \Big] \Big\},
\end{split}
\end{equation}
or using the recurrence relation for the Bessel functions,
\begin{equation}
    sx J'_\nu(x)=-\nu J_\nu(x) +xJ_{\nu-s}(x),\qquad s=\pm1,
\end{equation}
as
\begin{equation}\label{totAmpl1-2}
\begin{split}
	\mathcal{A} = &\,2\pi \sum_{\{n_i\},\{r_i\},\{p_{ij}\},\{q_{ij}\}=-\infty}^\infty
    \de_N\big(k_0(1-\tilde{n}_3\be_3)-\small{\sum_i} n_i\omega_i \big)e^{ik_3z_0}
    j_{m-\tiny{\sum_i} (n_i+2r_i) + 2\tiny{\sum_{ij}} p_{ij}}^0 \times\\
	&\times \prod_{i=1}^{M} \Big[ J_{n_i+r_i-\tilde{q}_i-\tilde{p}_i}(\rho_i) J_{r_i}(\de_i) e^{in_i\chi_i}\Big]
    \Big\{ \be_3 +\sum_{j=1}^M(1-\tilde{n}_3)\frac{\omega_j(n_j-\tilde{q}_j- \tilde{p}_j)}{k_\perp n_\perp} -\\
    &-
    \sum_{j=1}^{M}\frac{\omega_j}{k_\perp n_\perp} \Big(\frac{\rho_jJ_{n_j+r_j-\tilde{q}_j-\tilde{p}_j-s}(\rho_j)}{J_{n_j+r_j-\tilde{q}_j-\tilde{p}_j}(\rho_j)} -\frac{\de_jJ_{r_j-s}(\de_j)}{J_{r_j}(\de_j)} \Big)  \Big\}\prod_{k,l=1}^M\Big[J_{q_{kl}}(\De_{kl}) J_{p_{kl}}(\vk_{kl}) \Big].
\end{split}
\end{equation}
Since $\be_\perp\ll1$ and $n^2_\perp\ll1$, the last term on the second line can be neglected near the radiation harmonics \eqref{enSpect}. Indeed, it is easy to see that this term is approximately equal to
\begin{equation}
    \Big(1-\frac{\sum_j \omega_j (\tilde{q}_j+\tilde{p}_j)}{\sum_i\omega_i n_i}\Big) \frac{n_\perp^2\ga^2+\be_\perp^2\ga^2}{4\ga^2}.
\end{equation}

If $|\tilde{p}_i|\lesssim\sum_j|\vk_{ij}|\ll\ga^2$ and $|\tilde{q}_i|\lesssim\sum_j|\De_{ij}|\ll\ga^2$, then, for the frequencies $\omega_i$ that do not differ greatly, this term is much smaller than the contribution from $\be_3\approx1$. In this case, neglecting the small contributions, we have
\begin{equation}\label{totAmpl2}
\begin{split}
	\mathcal{A}\approx&\,  2\pi \sum_{\{n_i\},\{r_i\},\{p_{ij}\},\{q_{ij}\}=-\infty}^\infty
    \de_N\big(k_0(1-\tilde{n}_3\be_3)-\small{\sum_i} n_i\omega_i \big)e^{ik_3z_0}
    j_{m-\tiny{\sum_i} (n_i+2r_i) + 2\tiny{\sum_{ij}} p_{ij}}^0\times\\
	&\times  \prod_{i=1}^{M} \Big[ J_{n_i+r_i-\tilde{q}_i-\tilde{p}_i}(\rho_i) J_{r_i}(\de_i) e^{in_i\chi_i}\Big]
    \prod_{k,l=1}^M\Big[J_{q_{kl}}(\De_{kl}) J_{p_{kl}}(\vk_{kl}) \Big]\times\\
    &\times \Big\{ \be_3 -
    \sum_{j=1}^{M}\frac{\omega_j}{k_\perp n_\perp} \Big(\frac{\rho_jJ_{n_j+r_j-\tilde{q}_j-\tilde{p}_j-s}(\rho_j)}{J_{n_j+r_j-\tilde{q}_j-\tilde{p}_j}(\rho_j)} -\frac{\de_jJ_{r_j-s}(\de_j)}{J_{r_j}(\de_j)} \Big) \Big\}.
\end{split}
\end{equation}
Recall some notations used in this formula. The function $\de_N(x)$ is the delta-shaped sequence \eqref{delta-shaped_seq}, $\omega_i$ are the undulator frequencies, $\tilde{n}_3=\sqrt{1-n_\perp^2}$ and $n_\perp=k_\perp/k_0$, the constant $z_0$ is given in \eqref{trajectory}, the quantities $j^0_m$, $\rho_i$, $\de_i$, $\vk_{ij}$, and $\De_{ij}$ are defined in \eqref{a_b}, \eqref{c_d}, \eqref{R_D_x0}, and \eqref{j0_rho_delta_vk_Delta}, the undulator phases $\chi_i$ are specified after formula \eqref{magnetic_field}, $\be_3$ is the average particle velocity along the $z$ axis and is presented in \eqref{dip_parameter}.

It is also not difficult to see, taking into account the expressions for the oscillation amplitudes \eqref{c_d} and their connection to the undulator strength parameters \eqref{dip_parameter}, that, on substituting into the expression for the average number of radiated photons \eqref{probab_I}, the second term in $\dot{x}_3$ in \eqref{dotxpm} leads to a correction of the order or less than $n_\perp^2$ compared to the other terms in \eqref{probab_I}. Therefore, this contribution can be neglected for $n_\perp^2\ll1$, i.e., in the domain where the main part of radiation is concentrated.

In order to derive the probability of radiation of twisted photons, it is necessary to take modulus squared of expression \eqref{totAmpl1-2} or \eqref{totAmpl2} and substitute the result into formula \eqref{probab_I}. In this case, unlike a one-frequency undulator, the interference contributions arise from the different sets of $n_i$ for arbitrary ratios between the frequencies $\omega_i$. The expression for the average number of twisted photons is rather cumbersome and we will not write it out here.

In the case where $k_\perp|x^0_+|\ll1$, i.e., when the center of the ellipse of the spiral along which the particle moves is close to the axis relative to which the angular momentum of the twisted photons is defined, the following relation holds
\begin{equation}\label{center}
	j_{m-\tiny{\sum_i} (n_i+2r_i) + 2\tiny{\sum_{ij}} p_{ij}}^0\approx\de_{m,\tiny{\sum_i} (n_i+2r_i) - 2\tiny{\sum_{ij}} p_{ij}}.
\end{equation}
Then we obtain the selection rule for the radiation of twisted photons
\begin{equation}\label{sel_rule_gen}
    m+n_1+\cdots +n_M \text{ is an even number.}
\end{equation}
This rule is a generalization of the selection rule for radiation of twisted photons by an elliptical one-frequency undulator \cite{KazRyakElUnd}. Notice that the second term in the expression for $\dot{x}_3$ in \eqref{dotxpm} does not violate this selection rule. It should also be stressed that since the same energy value \eqref{enSpect} can be realized by different sets of $n_i$, this selection rule cannot always be observed directly in the radiation. Taking this selection rule into account, the average number of twisted photons \eqref{probab_I} radiated by an elliptical $M$-frequency undulator possesses a symmetry property: if one changes the chirality of undulator, i.e., for all $i$ and $j$,
\begin{equation}\label{refl_sym_1}
    \omega_i\rightarrow -\omega_i,\qquad \chi_i\rightarrow -\chi_i,\qquad \vk_{ij}\rightarrow-\vk_{ij},\qquad\De_{ij}\rightarrow-\De_{ij},
\end{equation}
then
\begin{equation}\label{refl_sym_2}
    dP(s,m,k_3,k_\perp)\rightarrow dP(-s,-m,k_3,k_\perp).
\end{equation}
This symmetry property follows from the fact that the amplitude \eqref{totAmpl1-2} changes by an insignificant phase $(-1)^m$ under the transformation \eqref{refl_sym_1} and $s\rightarrow-s$, $m\rightarrow-m$. Moreover, when \eqref{sel_rule_gen} holds true, the average number of radiated twisted photons does not change under shifting of all $\chi_i$ as $\chi_i\rightarrow\chi_i+\pi$.

\section{Two-frequency undulator}\label{Two-frequency_Undulator}

In this section, we consider in detail two-frequency undulators, i.e., we investigate the case $ M = 2 $. From general formula \eqref{totAmpl1-2}, we derive the one-particle radiation amplitude (see the notation in the previous sections, in particular, formulas \eqref{a_b}, \eqref{c_d},\eqref{R_D_x0}, and \eqref{j0_rho_delta_vk_Delta})
\begin{equation}\label{2omegaAmplGen}
\begin{split}
	\mathcal{A} = &\, 2\pi \sum_{\substack{n_1,n_2,r_1,r_1,p_{11},p_{12}\\p_{21},p_{22},q_{12},q_{21}=-\infty}}^\infty
    \de_N\big(k_0(1-\tilde{n}_3\be_3)- n_1\omega_1 -n_2\omega_2 \big)J_{q_{12}}(\De_{12})J_{q_{21}}(\De_{21}) \times \\
	&\times e^{ik_3z_0}j_{m- \tiny{\sum_i}(n_i+2r_i) + 2\tiny{\sum_{ij}}p_{ij}}^0
    \prod_{i=1}^{2} \Big[ J_{n_i+r_i-\tilde{q}_i-\tilde{p}_i}(\rho_i) J_{r_i}(\de_i)
    e^{in_i\chi_i}\Big] \prod_{k,l=1}^2\Big[J_{p_{kl}}(\vk_{kl})\Big]\times\\
    &\times\Big\{ \be_3 +\sum_{j=1}^2\Big[(1-\tilde{n}_3)\frac{\omega_j(n_j-\tilde{q}_j-\tilde{p}_j)}{k_\perp n_\perp}
    -
    \frac{\omega_j}{k_\perp n_\perp} \Big(\frac{\rho_jJ_{n_j+r_j -\tilde{q}_j-\tilde{p}_j-s}(\rho_j)}{J_{n_j+r_j-\tilde{q}_j-\tilde{p}_j}(\rho_j)} -\frac{\de_jJ_{r_j-s}(\de_j)}{J_{r_j}(\de_j)} \Big)\Big]  \Big\}.
\end{split}
\end{equation}
In this case (see formula \eqref{tilde_q_tilde_p}),
\begin{equation}
    \tilde{q}_2=-\tilde{q}_1=q_{21}-q_{12}.
\end{equation}
It is clear that the radiation energy spectrum \eqref{enSpect} has the form
\begin{equation}\label{enSpect2}
	k_0=\frac{n_1\omega_1+ n_2\omega_2}{1-\tilde{n}_3\be_3}=n_1\tilde{\omega}_1+n_2\tilde{\omega}_2,
\end{equation}
or
\begin{equation}
	k_0=\frac{\omega_1(n_1 + \eta_2 n_2)}{1-\tilde{n}_3\be_3}.
\end{equation}
If $\eta_2$ is a rational number and $\eta_2=h_2/g_2$, then (see formula \eqref{enSpect1}) $g=g_2$, $(\tilde{g}_1,\tilde{g}_2)=(g_2,h_2)$, and $d=1$. Consequently, in this case
\begin{equation}\label{enSpect2_1}
    k_0=\frac{\omega n}{1-\tilde{n}_3\be_3},\qquad \omega=\omega_1/g_2,
\end{equation}
and $(\la_1,\la_2)=(g_2,h_2)$.

\subsection{Helical two-frequency undulator}

Let us consider a helical two-frequency undulator. We set (see the expression for a particle trajectory \eqref{trajectory})
\begin{equation}
	a_1 = b_1 = r_1, \qquad a_2 = b_2 = r_2,
\end{equation}
and the magnetic field \eqref{magnetic_field} is to be
\begin{equation}
    H_x=H_x^1\sin \tilde{\vf}_1 + H_x^2\sin \tilde{\vf}_2,\qquad H_y=- H_x^1\cos \tilde{\vf}_1 -H_x^2\cos \tilde{\vf}_2,\qquad H_z=0.
\end{equation}
The magnetic field in this undulator is a linear superposition of magnetic fields of two one-frequency helical undulators. Then the quantities \eqref{R_D_x0} entering into the arguments of the Bessel functions become
\begin{equation}
	R_{1,2} = r_{1,2} , \qquad D_{1,2}= \de_{1,2}  = 0,
\end{equation}
and it follows from \eqref{c_d}, \eqref{j0_rho_delta_vk_Delta} that $\vk_{ij}=0$. Consequently, all the Bessel functions whose arguments contain $\de_{i}$ or $\vk_{ij}$ are replaced by the Kronecker symbols. As a result, we have from \eqref{2omegaAmplGen} that
\begin{multline}\label{2omegaAmpl1}
	\mathcal{A} =  2\pi \sum_{n_1,n_2=-\infty}^\infty
    \de_N\big(k_0(1-\tilde{n}_3\be_3)- n_1\omega_1 -n_2\omega_2 \big)
	 j_{m- n_1 -n_2 }^0 e^{ik_3z_0+in_1\chi_1+in_2\chi_2}\times\\
    \times\sum_{q_{12},q_{21}=-\infty}^\infty \Big[ \Big(\be_3 + (1-\tilde{n}_3)\frac{\omega_1 (n_1-\tilde{q}_1) +\omega_2 (n_2+\tilde{q}_1)}{k_\perp n_\perp}\Big)J_{n_1-\tilde{q}_1}(\rho_1) J_{n_2+\tilde{q}_1}(\rho_2)-\\
    -
    \frac{\omega_1\rho_1}{k_\perp n_\perp} J_{n_1-\tilde{q}_1-s}(\rho_1)J_{n_2+\tilde{q}_1}(\rho_2) -\frac{\omega_2\rho_2}{k_\perp n_\perp} J_{n_1-\tilde{q}_1}(\rho_1)J_{n_2+\tilde{q}_1-s}(\rho_2) \Big] J_{q_{12}}(\De_{12})J_{q_{21}}(\De_{21}).
\end{multline}
Let us show that the second term in round brackets on the second line can be neglected in the ultrarelativistic limit in the region where the main part of radiation is concentrated. In this case we have $\be_\perp\ll1$, condition \eqref{enSpect2} is satisfied, and $n_\perp\ga\lesssim\sqrt{1+K^2}$. Then
\begin{equation}
    |\De_{ij}|=\frac{k_3}{2\ga^2}\frac{K_iK_j}{|\omega_j-\omega_i|}\approx\frac{\sum_k\omega_k n_k}{|\omega_i-\omega_j|}\frac{K_iK_j}{K^2} \frac{K^2}{1+K^2+n_\perp^2\ga^2}<\frac{\sum_k\omega_k n_k}{|\omega_i-\omega_j|}\frac{K_iK_j}{K^2}.
\end{equation}
Supposing that
\begin{equation}
    \frac{\sum_k\omega_k n_k}{|\omega_i-\omega_j|}\frac{K_iK_j}{K^2}\ll\ga^2,
\end{equation}
and taking into account the properties of the Bessel functions, we obtain the estimate for the summation indices
\begin{equation}
    |q_{ij}|\lesssim|\De_{ij}|\ll\ga^2.
\end{equation}
Therefore, $|\tilde{q}_i|\ll\ga^2$ and so we can neglect the second term in round brackets on the second line of \eqref{2omegaAmpl1}. As a result, we arrive at the approximate expression
\begin{equation}\label{2omegaAmpl2}
\begin{split}
	\mathcal{A} \approx &\,  2\pi \sum_{n_1,n_2=-\infty}^\infty
    \de_N\big(k_0(1-\tilde{n}_3\be_3)- n_1\omega_1 -n_2\omega_2 \big)
	 j_{m- n_1 -n_2 }^0 e^{ik_3z_0+in_1\chi_1+in_2\chi_2}\times\\
    &\times \sum_{q_{12},q_{21}=-\infty}^\infty \Big[\be_3 J_{n_1-\tilde{q}_1}(\rho_1) J_{n_2+\tilde{q}_1}(\rho_2)
    -
    \frac{\omega_1\rho_1}{k_\perp n_\perp} J_{n_1-\tilde{q}_1-s}(\rho_1)J_{n_2+\tilde{q}_1}(\rho_2)-\\
    &-\frac{\omega_2\rho_2}{k_\perp n_\perp} J_{n_1-\tilde{q}_1}(\rho_1)J_{n_2+\tilde{q}_1-s}(\rho_2) \Big]  J_{q_{12}}(\De_{12})J_{q_{21}}(\De_{21}).
\end{split}
\end{equation}
This expression coincides with the amplitude of radiation of a twisted photon by an electron in a helical one-frequency undulator \cite{BKL2,BKL4,Epp2023,Pavlov2024} if we put $\rho_2 = 0$ and $\De_{12}=-\De_{21}=0$. For $ k_\perp |x_+^0| \ll 1 $, the selection rule arises,
\begin{equation}\label{sel_rul2}
    m = n_1+n_2,
\end{equation}
which is a generalization of the selection rule for the radiation of twisted photons by a helical one-frequency undulator \cite{SasMcNu,HemMarRos11,HemMar12,HKDXMHR,BHKMSS,Rubic17,BKL2,BKL4}. This selection rule and the radiation energy spectrum \eqref{enSpect2} have a simple interpretation in terms of photons emitted by the electron and exchanged with the undulator. In moving through the undulator, the electron absorbs from the undulator (when $n_i\omega_i>0$) or gives back to it (when $n_i\omega_i<0$) $|n_i|$ virtual photons with frequency $|\omega_i|$ for each $i=\overline{1,2}$, and also emits a single real photon with frequency \eqref{enSpect2}. The virtual photons possess a helicity $\sgn\omega_i$. Because $\sgn n_i=\sgn \omega_i$ when the electron absorbs a virtual photon created by the undulator, and $\sgn n_i=-\sgn\omega_i$ when the virtual photon is transferred from the electron to the undulator, the selection rule \eqref{sel_rul2} just expresses the conservation law for the projection of TAM onto the $z$ axis, i.e., the projection of TAM got by the electron from the undulator is transferred to the radiated photon.

\begin{figure}[t!]
\centering
i)\;\,\raisebox{-0.5\height}{\includegraphics*[width=0.750\linewidth]{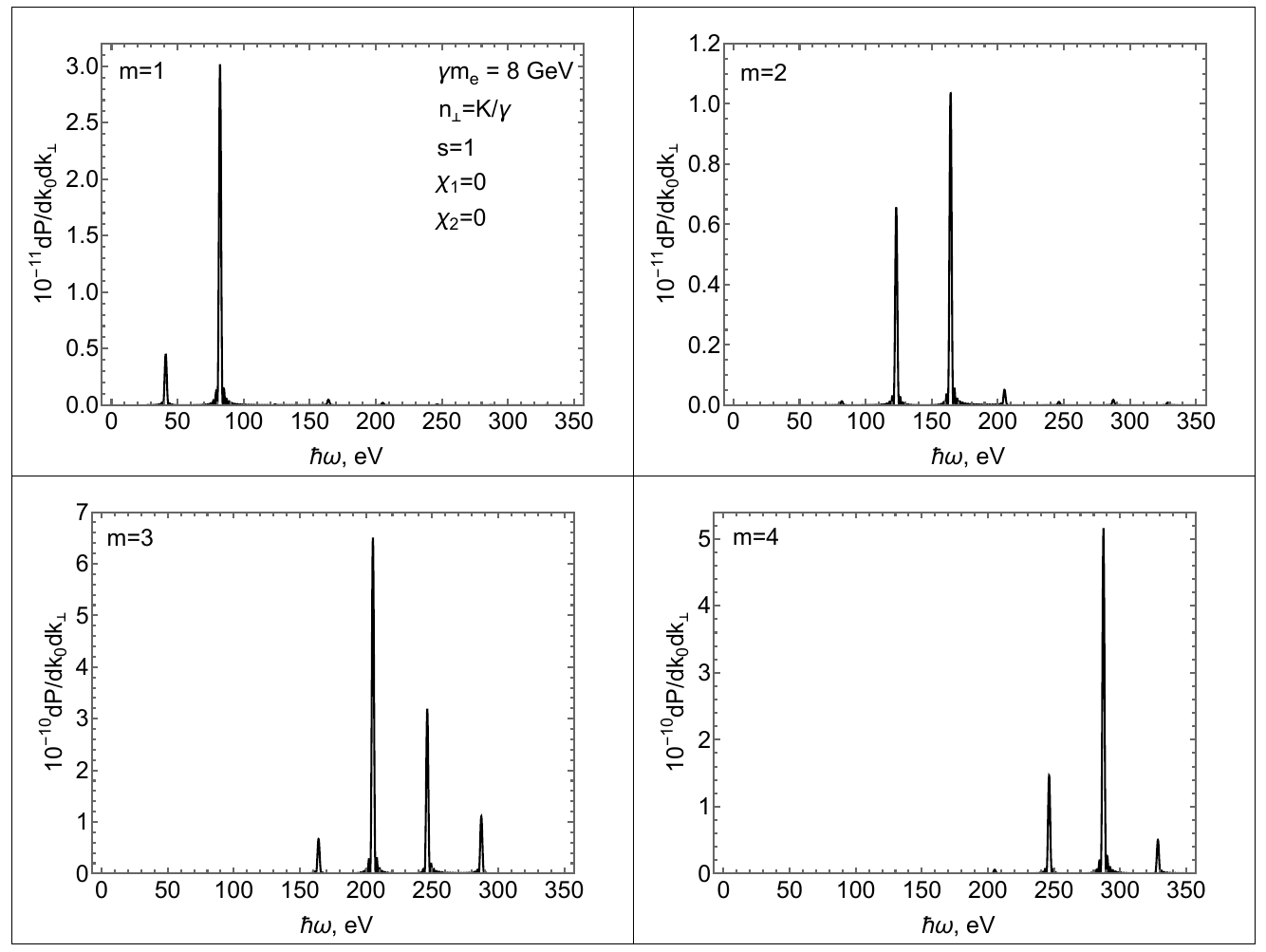}}\\
ii)\;\raisebox{-0.5\height}{\includegraphics*[width=0.746\linewidth]{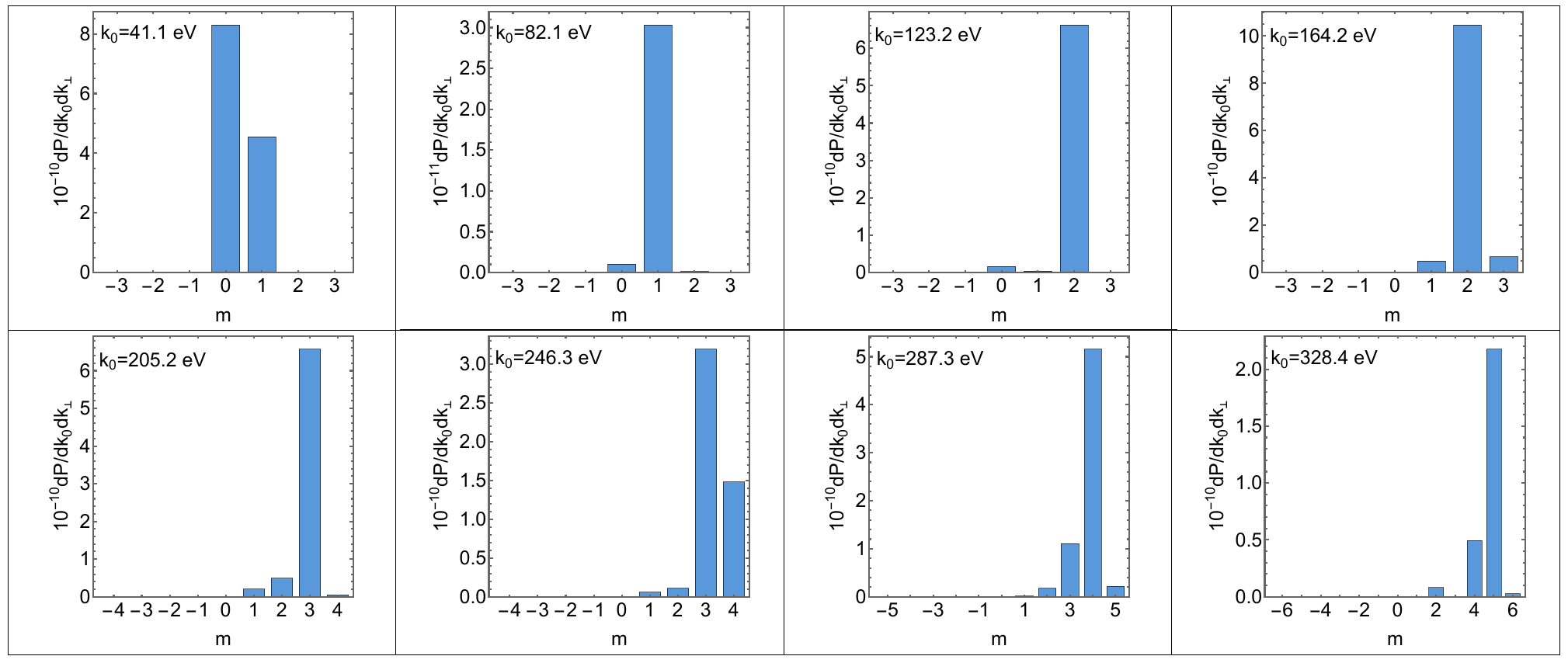}}
\caption{{\footnotesize The energy and TAM projection spectra of twisted photon radiation from the helical two-frequency undulator. The Lorentz factor of electrons $\ga=1.566\times 10^4$, the magnetic field strengths in undulator $H_x^i=-H_y^i=1.16\times 10^4$ G, $i=\overline{1,2}$, and the number of undulator sections $N=40$. The frequencies of subundulators are $\omega_1=2.07\times 10^{-5}$ eV, $\omega_2=3.10\times 10^{-5}$ eV and so the undulator strength parameters are $K_1=6.5$, $K_2=4.3$, $K=7.8$, and the frequencies $\tilde{\omega}_1=82.1$ eV, $\tilde{\omega}_2=123.2$ eV. Therefore, $\eta_2=3/2$, $\la_1=2$, $\la_2=3$, and the respective B\'{e}zout coefficients become $n_1^0=2$, $n_2^0=-1$. It is clear from the plots (ii) that the selection rule \eqref{sel_rul2-1} is fulfilled. The restrictions on the numbers of virtual photons, $|n_i|$, discussed after Eqs. \eqref{n_i_restrictions}, \eqref{q_ij_restrictions} determine the positions of the main peaks in the distribution over $m$.}}
\label{Plot_2helical32}
\end{figure}

\begin{figure}[t!]
\centering
i)\;\raisebox{-0.5\height}{\includegraphics*[width=0.745\linewidth]{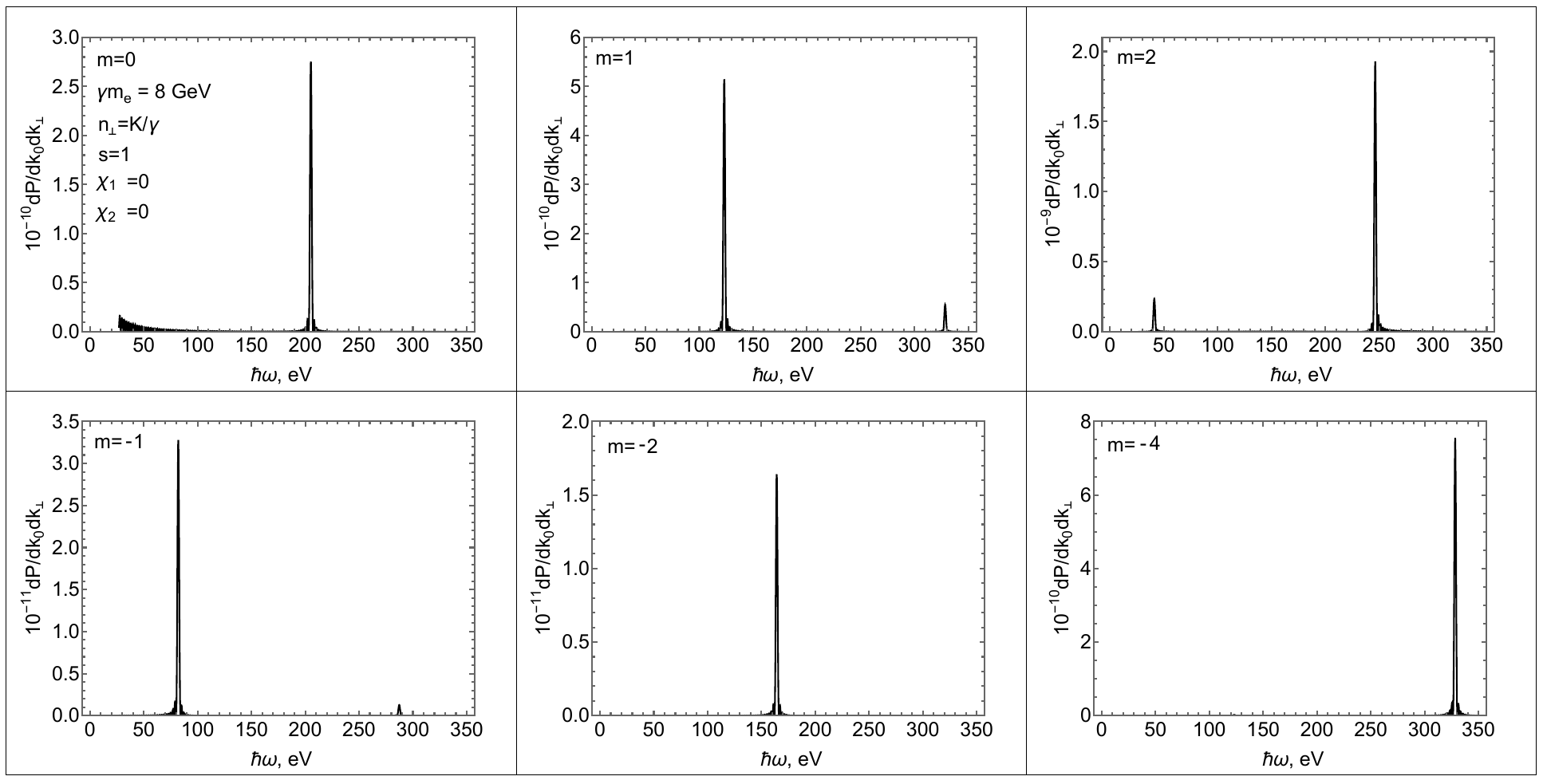}}\\
ii)\;\raisebox{-0.5\height}{\includegraphics*[width=0.748\linewidth]{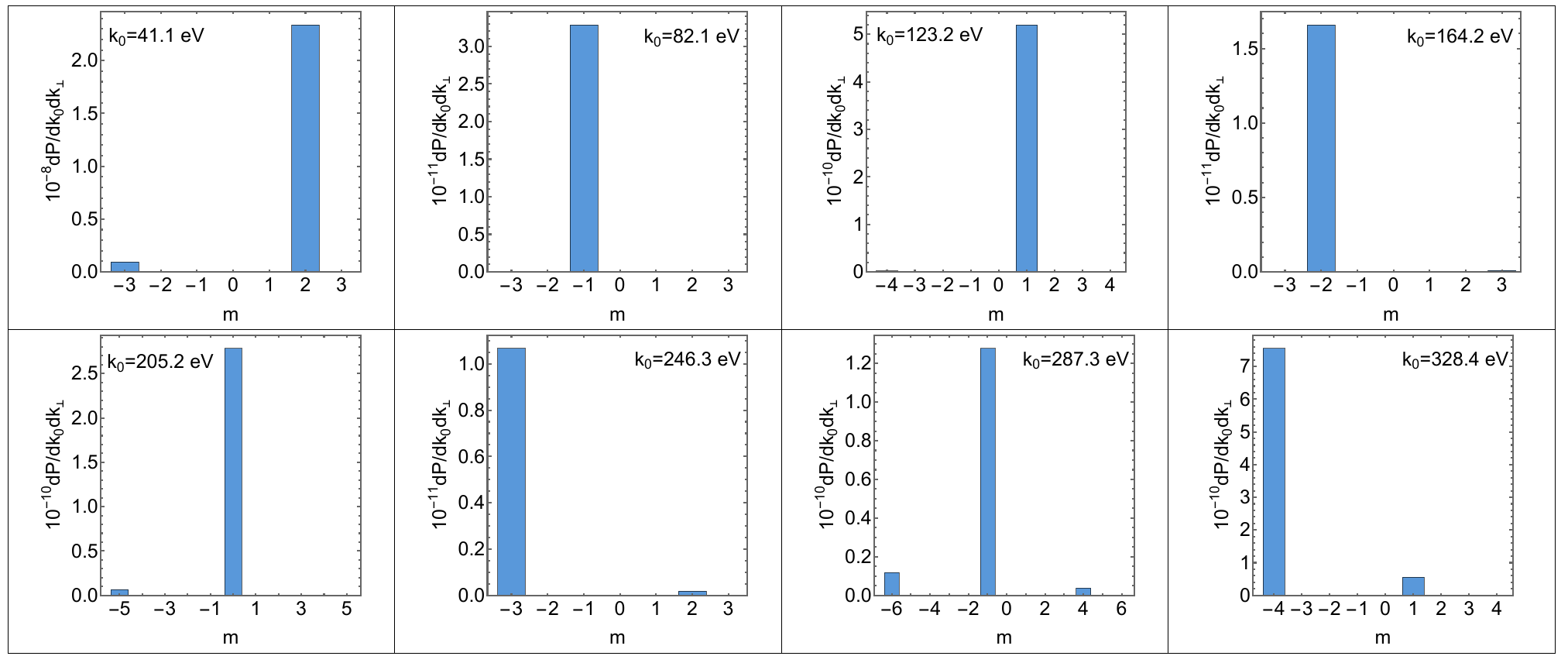}}
\caption{{\footnotesize The energy and TAM projection spectra of twisted photon radiation from the helical two-frequency undulator. The Lorentz factor of electrons $\ga=1.566\times 10^4$, the magnetic field strengths in undulator $H_x^i=-H_y^i=1.16\times 10^4$ G, $i=\overline{1,2}$, and the number of undulator sections $N=40$. The frequencies of subundulators are $\omega_1=-2.07\times 10^{-5}$ eV, $\omega_2=3.10\times 10^{-5}$ eV and so the undulator strength parameters are $K_1=6.5$, $K_2=4.3$, $K=7.8$, and the frequencies $\tilde{\omega}_1=-82.1$ eV, $\tilde{\omega}_2=123.2$ eV. Therefore, $\eta_2=-3/2$, $\la_1=2$, $\la_2=-3$, and the respective B\'{e}zout coefficients become $n_1^0=-1$, $n_2^0=-1$. It is clear from the plots (ii) that the selection rule \eqref{sel_rul2-1} is fulfilled. The restrictions on the numbers of virtual photons, $|n_i|$, discussed after Eqs. \eqref{n_i_restrictions}, \eqref{q_ij_restrictions} determine the positions of the main peaks in the distribution over $m$. The plot (i) with $m=0$ is depicted only for the photon energies $k_0>27$ eV to cut off the large infrared contribution of edge radiation.}}
\label{Plot_2helical32m}
\end{figure}

Let us find the spectrum with respect to the TAM projection $m$ at a fixed energy \eqref{enSpect2} in the case when the ration of undulator frequencies, $\eta_2$, is a rational number. From \eqref{enSpect1} and \eqref{enSpect2_1}, we have
\begin{equation}\label{Dioph_eqn}
    n_1\la_1+n_2\la_2=n\in \Z,\qquad n\neq0.
\end{equation}
As has been discussed in detail in the previous section, $\sgn n=\sgn\omega=\sgn\omega_1$. On solving equation \eqref{Dioph_eqn}, we obtain
\begin{equation}\label{Dioph_eqn_sol}
    (n_1,n_2)=n(n_1^0,n_2^0)+(-\la_2,\la_1)k,\qquad k\in\Z,
\end{equation}
where $(n_1^0,n_2^0)$ are the B\'{e}zout coefficients for the pair of integers $(\la_1,\la_2)$. Substituting this solution into \eqref{sel_rul2}, we arrive at the selection rule for the TAM projection
\begin{equation}\label{sel_rul2-1}
    m=n(n_1^0+n_2^0) +(\la_1-\la_2)k.
\end{equation}
This selection rule allows us to find the admissible values of $m$ for a fixed principal quantum number $n$. It is clear from this selection rule that, for a helical two-frequency undulator composed of helical one-frequency undulators of opposite chirality ($\la_1$ and $\la_2$ are of different signs), the interval between the values of $m$ realized on the radiation harmonics is greater than for a helical two-frequency undulator composed of the same one-frequency undulators of the same chirality ($\la_1$ and $\la_2$ of the same sign). A numerical verification of the selection rule \eqref{sel_rul2-1} is shown in Figs. \ref{Plot_2helical32}, \ref{Plot_2helical32m}.

The solution \eqref{Dioph_eqn_sol} of the Diophantine equation \eqref{Dioph_eqn} allows us to cast the amplitude of the twisted photon radiation \eqref{2omegaAmpl2} into the form
\begin{equation}\label{2omegaAmpl_simpl}
\begin{split}
	\mathcal{A} = &\,  2\pi \sum_{n=\sgn(\omega)}^{\sgn(\omega)\infty}
    \de_N\big(k_0(1-\tilde{n}_3\be_3)- \omega n \big)
	 \sum_{k=-\infty}^\infty j_{m- n_1 -n_2 }^0 e^{ik_3z_0+in_1\chi_1+in_2\chi_2}\times\\
    &\times \sum_{q_{12},q_{21}=-\infty}^\infty \Big[\be_3 J_{n_1-\tilde{q}_1}(\rho_1) J_{n_2+\tilde{q}_1}(\rho_2)
    -
    \frac{\omega_1\rho_1}{k_\perp n_\perp} J_{n_1-\tilde{q}_1-s}(\rho_1)J_{n_2+\tilde{q}_1}(\rho_2)-\\
    &-\frac{\omega_2\rho_2}{k_\perp n_\perp} J_{n_1-\tilde{q}_1}(\rho_1)J_{n_2+\tilde{q}_1-s}(\rho_2) \Big]  J_{q_{12}}(\De_{12})J_{q_{21}}(\De_{21}).
\end{split}
\end{equation}
where it is assumed that $n_{1,2}$ have the form \eqref{Dioph_eqn_sol}. The radiation amplitude \eqref{2omegaAmpl_simpl} can be written more compactly by introducing a special function
\begin{equation}
\begin{split}
    J_{n_1n_2}(\rho_1,\rho_2,\De_{12}):=&\sum_{q_{12},q_{21}=-\infty}^\infty J_{n_1-q_{12}-q_{21}}(\rho_1) J_{n_2+q_{12}+q_{21}}(\rho_2) J_{q_{12}}(\De_{12})J_{q_{21}}(\De_{12})=\\
    =&\sum_{q=-\infty}^\infty J_{n_1-q}(\rho_1) J_{n_2+q}(\rho_2) J_{q}(2\De_{12}).
\end{split}
\end{equation}
Then
\begin{equation}\label{2omegaAmpl_simpl1}
\begin{split}
	\mathcal{A} = &\,  2\pi \sum_{n=\sgn(\omega)}^{\sgn(\omega)\infty}
    \de_N\big(k_0(1-\tilde{n}_3\be_3)- \omega n \big)
	 \sum_{k=-\infty}^\infty j_{m- n_1 -n_2 }^0 e^{ik_3z_0+in_1\chi_1+in_2\chi_2}\times\\
    &\times \Big[\be_3 J_{n_1n_2}(\rho_1,\rho_2,\De_{12})
    -
    \frac{\omega_1\rho_1}{k_\perp n_\perp} J_{n_1-s,n_2}(\rho_1,\rho_2,\De_{12})
    -\frac{\omega_2\rho_2}{k_\perp n_\perp} J_{n_1,n_2-s}(\rho_1,\rho_2,\De_{12}) \Big].
\end{split}
\end{equation}
Thus we see that the photons are generated at the $n$-th harmonic in a coherent superposition of states with the different projections $m$ and the phases
\begin{equation}
    (m-n_1-n_2)\arg(x_+^0)+n_1\chi_1+n_2\chi_2.
\end{equation}
We have discarded the phase $k_3z_0$, which is the same for all the modes. In particular, if $ k_\perp |x_+^0| \ll 1 $, then the selection rule \eqref{sel_rul2}, \eqref{sel_rul2-1} is satisfied, and the phase of each mode is equal to
\begin{equation}\label{rel_phase_circ}
    n_1\chi_1+n_2\chi_2.
\end{equation}
No more than one term is nonzero in the sum over $k$ in the amplitude \eqref{2omegaAmpl_simpl}, i.e., for given $n$ and $m$, the quantum numbers $n_1$ and $n_2$ are uniquely specified provided, of course, expression \eqref{sel_rul2-1} considered as an equation for $k$ has a solution. As we see, the frequency, the relative phase, and the amplitude of modes with different $m$ can be controlled by changing the parameters of the undulator $a_i$, $b_i$, $\omega_i$, and $\chi_i$. Such superpositions of states of twisted photons with different $m$ were used in the papers \cite{Stock2015,Sherwin2017,Sherwin2020,Ivanov2022} to study interference effects in the processes of quantum electrodynamics with twisted photons. The average number of radiated twisted photons \eqref{probab_I} with $ k_\perp |x_+^0| \ll 1 $ is given by
\begin{multline}\label{2omega_probab}
	dP(s,m,k_3,k_\perp) =  \al\pi \sum_{n=\sgn(\omega)}^{\sgn(\omega)\infty}
    \de^2_N\big(k_0(1-\tilde{n}_3\be_3)- \omega n \big)
	 \sum_{k=-\infty}^\infty \de_{m,n_1 +n_2 }\times\\
    \times\Big[\be_3 J_{n_1n_2}(\rho_1,\rho_2,\De_{12})
    -
    \frac{\omega_1\rho_1}{k_\perp n_\perp} J_{n_1-s,n_2}(\rho_1,\rho_2,\De_{12})
    -\frac{\omega_2\rho_2}{k_\perp n_\perp} J_{n_1,n_2-s}(\rho_1,\rho_2,\De_{12}) \Big]^2 n_\perp^3 dk_3 dk_\perp,
\end{multline}
where we have neglected the interference contributions of harmonics with different $n$ and as above it is assumed that $n_{1,2}$ have the form \eqref{Dioph_eqn_sol}.

Since, in increasing $|n|$, the Bessel function $J_n(\rho)$ tends rapidly to zero for $|n|\gtrsim |\rho|$, there exist restrictions on the quantum numbers $n_i$,
\begin{equation}\label{n_i_restrictions}
    |n_i-\tilde{q}_i|\lesssim \rho_i\approx \Big|n\frac{\omega}{\omega_i}\frac{K_i}{K}\Big| \frac{2 K n_\perp\ga}{1+K^2+n_\perp^2\ga^2}<\Big|n\frac{\omega}{\omega_i}\frac{K_i}{K}\Big|,
\end{equation}
and
\begin{equation}\label{q_ij_restrictions}
    |q_{ij}|\lesssim|\De_{ij}|\approx \frac{\omega n}{|\omega_i-\omega_j|} \frac{K_iK_j}{1+K^2+n_\perp^2\ga^2}= \frac{ |n|}{|\la_i-\la_j|}\frac{K_iK_j}{K^2} \frac{K^2}{1+K^2+n_\perp^2\ga^2} <|n|\frac{K_iK_j}{K^2}<|n|.
\end{equation}
Therefore, among all the sets $\{n_i\}$ corresponding to a given principal quantum number $n$, the combinations in which the number of virtual photons $|n_i|$ does not exceed the modulus of harmonic number $|n|$ by the the order of magnitude are most likely to be realized. Moreover, as follows from \eqref{n_i_restrictions}, such $n_i$ are more likely to be realized where the index $i$ corresponds to virtual photons with lower energies $|\omega_i|$ and larger amplitudes $K_i$. For these reasons, the radiation at harmonics with energies $k_0<\min|\tilde{\omega}_i|$ is suppressed as compared to the radiation at harmonics dominating in the dipole regime where only one of the quantum numbers $n_i$ is different from zero and equals $\sgn \omega_i$.

\begin{figure}[t!]
	\centering
	i)\;\,\raisebox{-0.5\height}{\includegraphics*[width=0.750\linewidth]{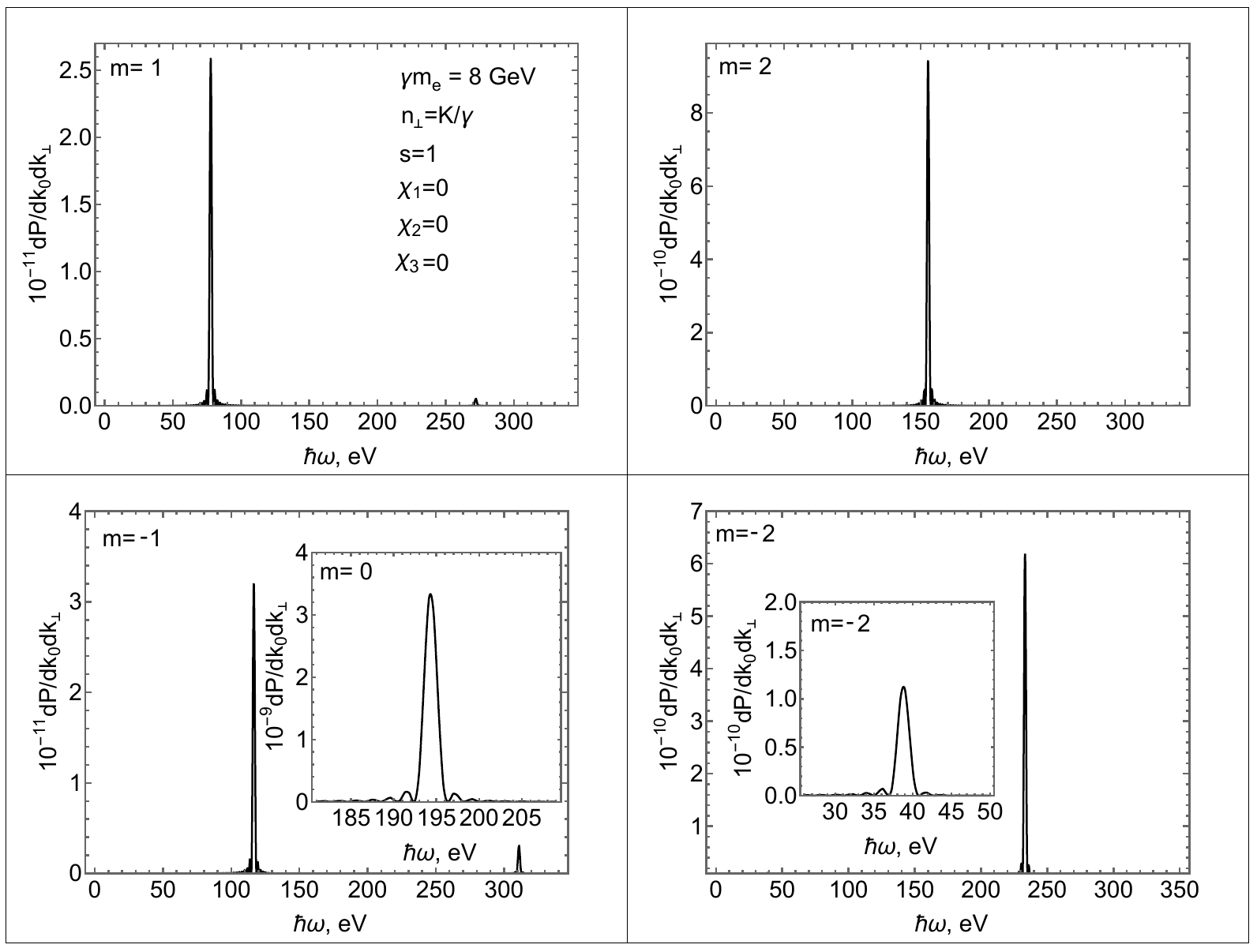}}\\
	ii)\;\raisebox{-0.5\height}{\includegraphics*[width=0.746\linewidth]{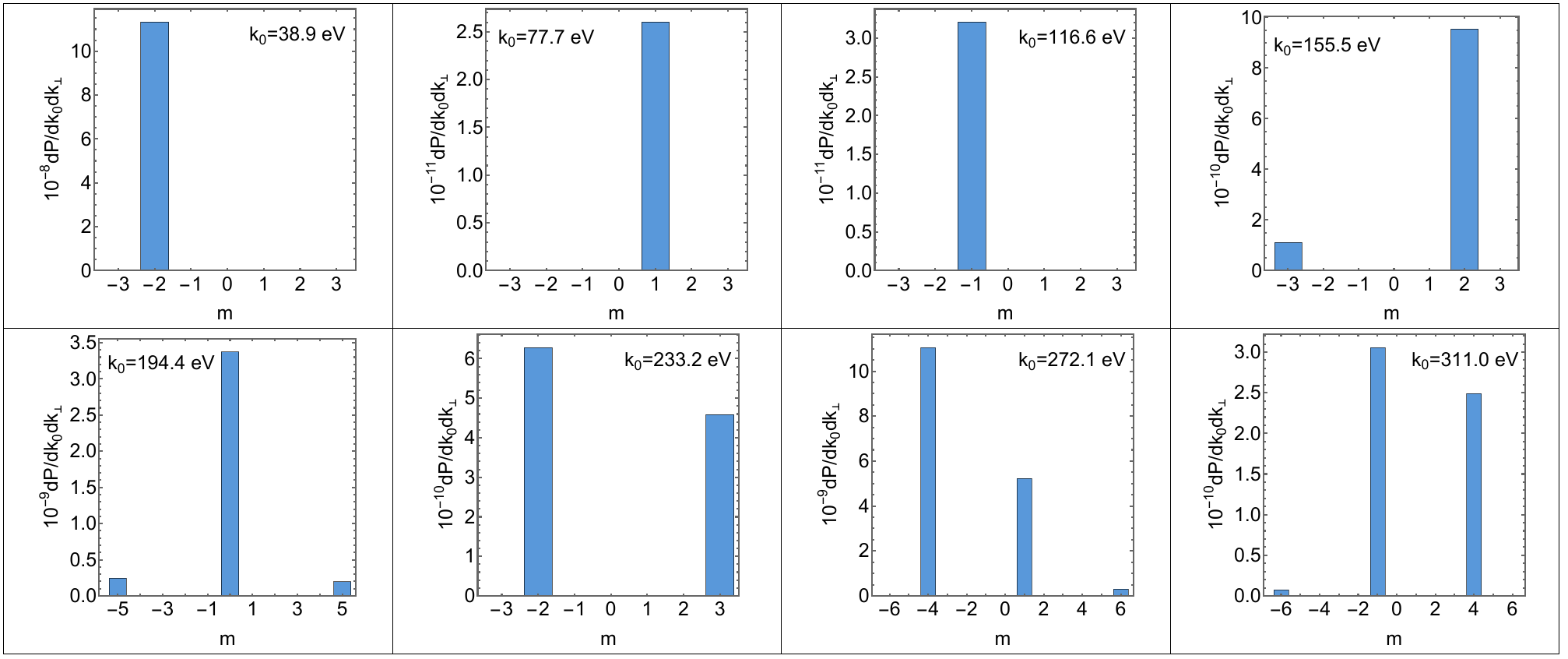}}
	\caption{{\footnotesize The energy and TAM projection spectra of twisted photon radiation from the helical three-frequency undulator. The Lorentz factor of electrons $\ga=1.566\times 10^4$, the magnetic field strengths in undulator $H_x^i=-H_y^i=1.16\times 10^4$ G, $i=\overline{1,2,3}$, and the number of undulator sections $N=40$. The frequencies of subundulators are $\omega_1=2.07\times 10^{-5}$ eV, $\omega_2=-3.09\times 10^{-5}$ eV, $\omega_3=7.23\times 10^{-5}$ eV and so the undulator strength parameters are $K_1=6.5$, $K_2=4.3$,  $K_3=1.9$, $K=8.04$, and the frequencies $\tilde{\omega}_1=77.7$ eV, $\tilde{\omega}_2=-116.6$ eV, $\tilde{\omega}_3=272.1$ eV. Therefore, $\eta_2=-3/2$, $\eta_3=7/2$, $\la_1=2$, $\la_2=-3$, $\la_3 = 7$ and the respective B\'{e}zout coefficients become $n_1^0=-1$, $n_2^0=-1$, $ n_3^0=0$. It is clear from the plots (ii) that the selection rule \eqref{sel_rul_circ} is fulfilled. The restrictions on the numbers of virtual photons, $|n_i|$, discussed after Eqs. \eqref{n_i_restrictions}, \eqref{q_ij_restrictions} determine the positions of the main peaks in the distribution over $m$. }}
	\label{Plot_3helical237}
\end{figure}

Let us describe the properties of radiation energy spectrum for a helical two-frequency undulator in the case when the one-frequency subundulator with frequency $\omega_2$ works in the dipole regime, i.e., $K_2\ll1$. In that case, $\rho_2\ll1$, $|\De_{12}|\ll1$, and so the main contributions to the radiation are given by the terms in \eqref{2omegaAmpl2} with $q_{12}=q_{21}=0$ and $n_2=\{-1,0,1\}$. In other words, no more than one virtual photon with frequency $|\omega_2|$ mediates between the electron and the undulator. Then the radiation spectrum of a helical two-frequency undulator looks as follows. There are the main radiation harmonics, $n_1\tilde{\omega}_1$, for which the selection rule $m=n_1$ is satisfied. Every such harmonic possesses the side harmonics with frequencies $n_1\tilde{\omega}_1\pm\tilde{\omega}_2$, where it is assumed that this expression is greater than zero. The intensity of the side harmonics is much less than that of the main harmonic and the radiation of twisted photons at these harmonics obeys the selection rule, $m=n_1\pm1$, respectively. There is also the harmonic with frequency $|\tilde{\omega}_2|$ for which $m=\sgn \omega_2$. The plots of the average number of twisted photons emitted in a helical two-frequency undulator are shown for various parameters in Figs. \ref{Plot_2helical32}, \ref{Plot_2helical32m}.

\begin{figure}[t!]
\centering
i)\;\,\raisebox{-0.5\height}{\includegraphics*[width=0.745\linewidth]{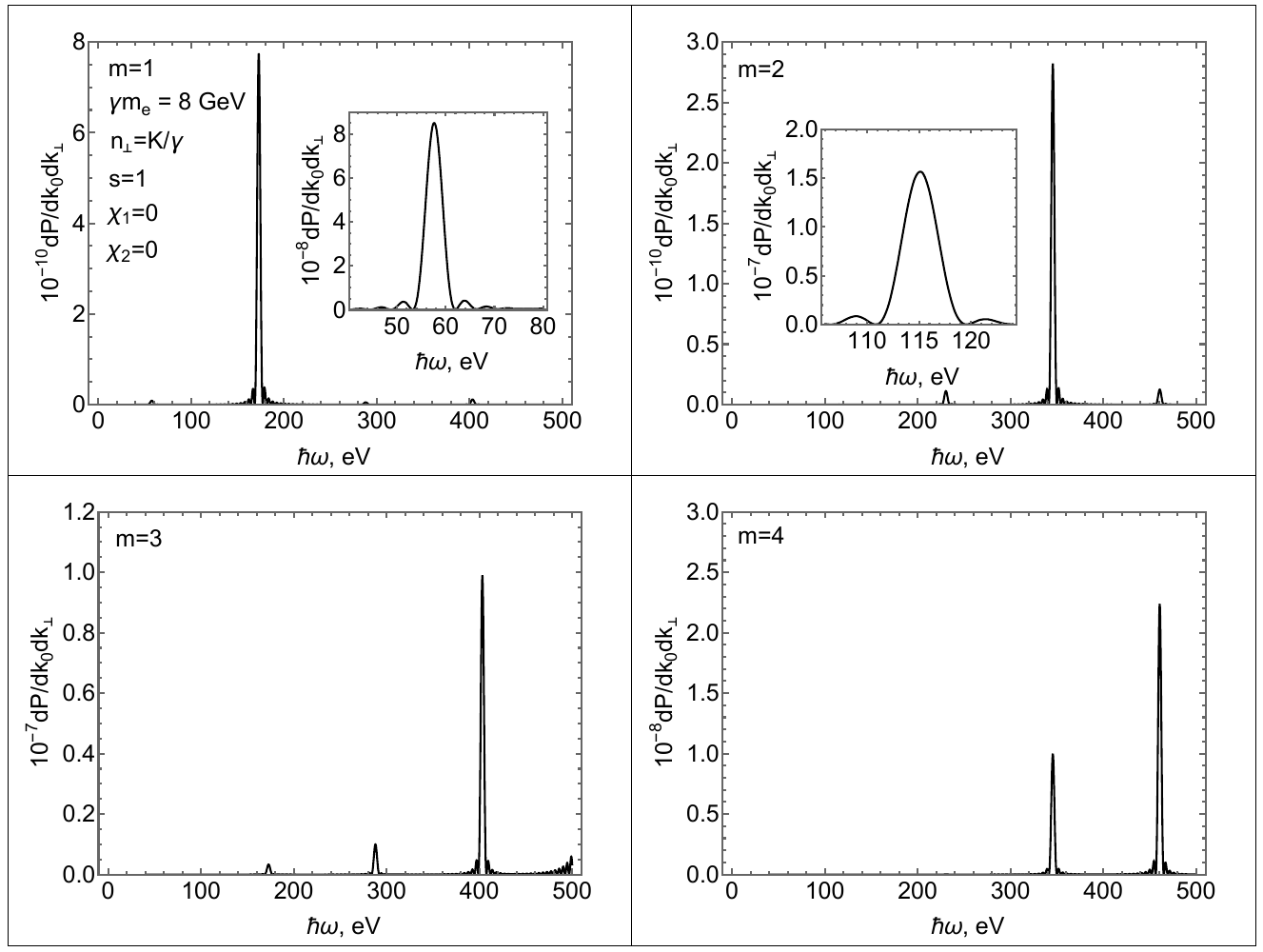}}\\
ii)\;\raisebox{-0.5\height}{\includegraphics*[width=0.744\linewidth]{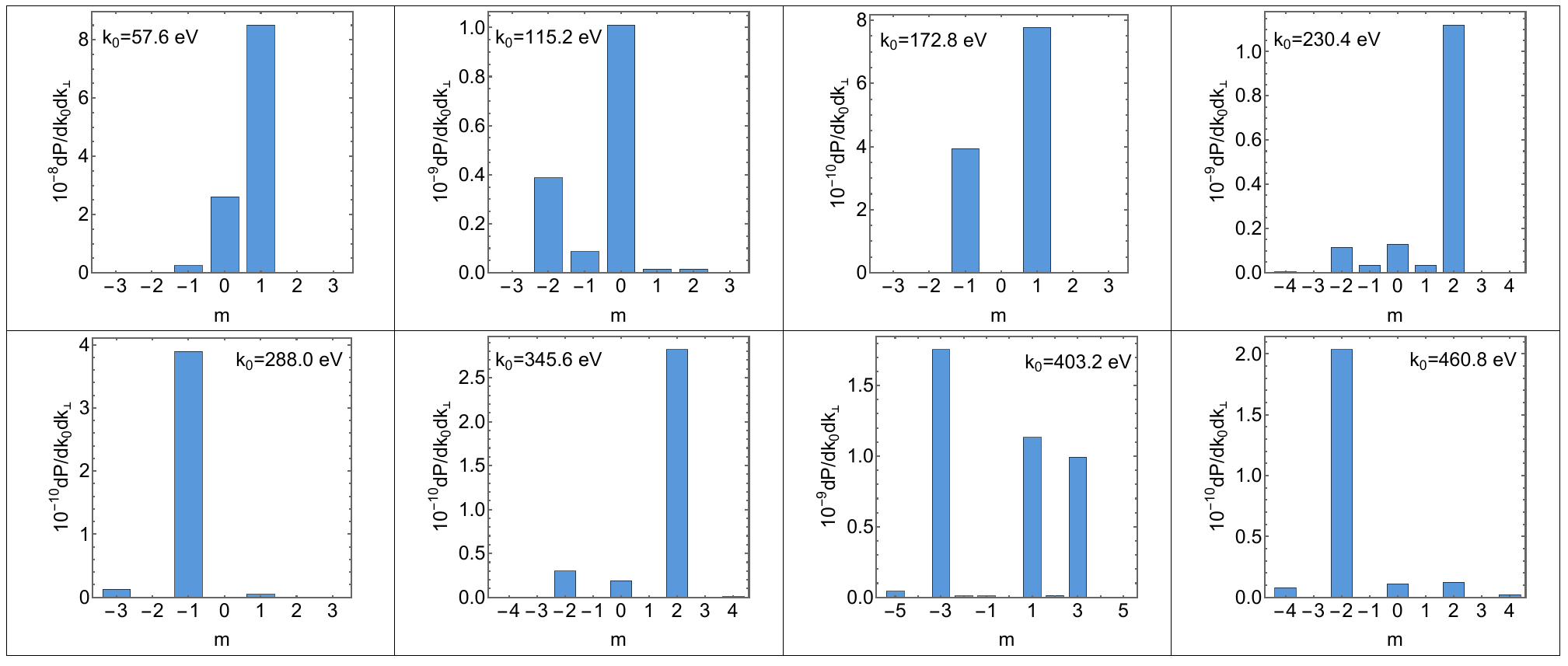}}
\caption{{\footnotesize The energy and TAM projection spectra of twisted photon radiation from the composition of two planar orthogonal undulators. The Lorentz factor of electrons $\ga=1.566\times 10^4$, the magnetic field strengths in undulator $H_x^2=-H_y^1=1.16\times 10^4$ G, $H_x^1=H_y^2=0$ G, and the number of undulator sections $N=40$. The frequencies of subundulators are $\omega_1=2.07\times 10^{-5}$ eV, $\omega_2=3.44\times 10^{-5}$ eV and so the undulator strength parameters are $K_1=4.6$, $K_2=2.8$, $K=5.4$, and the frequencies $\tilde{\omega}_1=172.8$ eV, $\tilde{\omega}_2=288.0$ eV. Therefore, $\eta_2=5/3$, $\la_1=3$, $\la_2=5$, and the respective B\'{e}zout coefficients become $n_1^0=2$, $n_2^0=-1$. It is clear from the plots (ii) that the selection rule \eqref{sel_rul_2planar1} is fulfilled. The contribution with $m=0$ at the harmonic $n=1$ that seems to violate the selection rule stems from the neighboring harmonic with $n=2$ and from edge radiation.}}
\label{Plot_2planar}
\end{figure}

It is not difficult to generalize the selection rules \eqref{sel_rul2-1} to the case of a helical $M$-frequency undulator with rational $\eta_i$ for $k_\perp|x_+^0|\ll1$. Equation \eqref{Dioph_eqn} in this case has the form
\begin{equation}\label{Dioph_eqn_M}
    n_1\la_1+n_2\la_2+\cdots+n_M\la_M=n\in \Z.
\end{equation}
The generalization of the selection rule \eqref{sel_rul2} becomes
\begin{equation}\label{sel_rulM}
    m = n_1+n_2+\cdots+n_M.
\end{equation}
The general solution to equation \eqref{Dioph_eqn_M} can be written in the form \cite{Quinlan2022}
\begin{equation}\label{Dioph_eqn_M_sol}
    n_i = nn_i^0 + \sum_{a>b=1}^{M} f_i^{(a,b)} k_{(a,b)},\qquad k_{(a,b)}\in\Z,
\end{equation}
where $n_i^0$ are the B\'{e}zout coefficients for the set of integers $(\la_1,\cdots,\la_M)$ and
\begin{equation}
    f^{(a,b)}_i=\la_{[a} \de_{b]i},
\end{equation}
where the square brackets denote antisymmetrization without the factor $1/2$. Notice that the representation \eqref{Dioph_eqn_M_sol} for the solution of equation \eqref{Dioph_eqn_M} for $M\geqslant3$ is not minimal in the sense that different $k_{(a,b)}$ may correspond to the same values of $n_i$. This shortcoming of representation \eqref{Dioph_eqn_M_sol} is not important for a further discussion. Substituting the solution \eqref{Dioph_eqn_M_sol} into \eqref{sel_rulM}, we obtain
\begin{equation}\label{sel_rul_circ}
    m=n\sum_{i=1}^M n_i^0 + \sum_{a>b=1}^{M} (\la_a-\la_b) k_{(a,b)},
\end{equation}
that generalizes the selection rule \eqref{sel_rul2-1} to the case of a helical $M$-frequency undulator. The interpretation of the selection rule \eqref{sel_rulM} and the radiation energy spectrum \eqref{enSpect} in terms of photons is the same as the one given above for the case of a helical two-frequency undulator. The restrictions \eqref{n_i_restrictions}, \eqref{q_ij_restrictions} on the number of virtual photons $|n_i|$ are left intact. The radiation spectrum in the case when one or more subundulators of a multifrequency undulator operate in the dipole regime is described in the same way as above for a helical two-frequency undulator. A numerical verification of the selection rule \eqref{sel_rul_circ} is shown in Fig. \ref{Plot_3helical237} for the three-frequency undulator.

\subsection{Composition of two planar undulators}

In this section, we consider the properties of the radiation of twisted photons by a two-frequency undulator consisting of two planar one-frequency undulators with different frequencies, whose planes are perpendicular to each other. To put it another way, we investigate the case when
\begin{equation}
    \rho_1=\de_1,\qquad \rho_2=-\de_2,
\end{equation}
which corresponds to $b_1=a_2=0$ (see formulas \eqref{R_D_x0}, \eqref{j0_rho_delta_vk_Delta}) and the magnetic field \eqref{magnetic_field} is given by
\begin{equation}\label{H_field_comp_two_planar}
    H_x= H_x^2\sin \tilde{\vf}_2,\qquad H_y= H_y^1\cos \tilde{\vf}_1,\qquad H_z=0.
\end{equation}
Just for reference, we relate this magnetic field to the magnetic field of two-frequency undulator investigated in \cite{Mirian2014}. Comparing \eqref{H_field_comp_two_planar} with formula (1) of \cite{Mirian2014}, we see that $H_y^2=0$, $H_x^2=B_{w2}$, $\chi_2=0$, $l_2=\la_{02}$, and $\omega_2=-2\pi\be_3/l_2$. Besides, $H_x^1=0$, $H_y^1=B_{w1}$, $\chi_1=\pi/2$, $l_1=\la_{01}$, and $\omega_1=-2\pi\be_3/l_1$. The undulator strength parameters $K_{1,2}$ coincide with that of \cite{Mirian2014} apart from the common factor $1/(2\pi\sqrt{2})$. The trajectory \eqref{trajectory} agrees with the trajectory given in formula (15) of \cite{Mirian2014} save some factors $2\pi$ missed in \cite{Mirian2014} in the definition of $K_{1,2}$.

In the case \eqref{H_field_comp_two_planar}, it follows from \eqref{c_d}, \eqref{j0_rho_delta_vk_Delta} that $\De_{ij}=0$ and therefore (see expression \eqref{2omegaAmplGen} for the radiation amplitude) $q_{12}=q_{21}=0$. In addition, $\vk_{12}=\vk_{21}=0$ (see formulas \eqref{c_d}, \eqref{j0_rho_delta_vk_Delta}) that implies $p_{12}=p_{21}=0$. Besides,
\begin{equation}
    \vk_{11}=\frac{k_3}{8}\omega_1a_1^2,\qquad \vk_{22}=-\frac{k_3}{8}\omega_2 b_2^2.
\end{equation}
Then we have approximately from formula \eqref{2omegaAmplGen} in the parameter space where the main part of radiation is concentrated that
\begin{multline}\label{2omegaAmpl_plane}
	\mathcal{A} \approx 2\pi \sum_{n_1,r_1,p_{11}=-\infty}^\infty \sum_{n_2,r_2,p_{22}=-\infty}^\infty
    \de_N\big(k_0(1-\tilde{n}_3\be_3)- n_1\omega_1 -n_2\omega_2 \big)\times \\
	\times (-1)^{r_2}e^{ik_3z_0}j_{m- n_1 -n_2 - 2 ( r_1+r_2 -p_{11} -p_{22})}^0 \prod_{i=1}^{2} \Big[ J_{n_i+r_i-2p_{ii}}(\rho_i)
    J_{r_i}(\rho_i)
    J_{p_{ii}}(\vk_{ii})e^{in_i\chi_i}\Big]\times\\
    \times \Big[ \be_3 -\frac{\omega_1\rho_1}{k_\perp n_\perp}\Big(\frac{J_{n_1+r_1-2p_{11}-s}(\rho_1)}{J_{n_1+r_1-2p_{11}}(\rho_1)} -\frac{J_{r_1-s}(\rho_1)}{J_{r_1}(\rho_1)} \Big) -\frac{\omega_2\rho_2}{k_\perp n_\perp}\Big(\frac{J_{n_2+r_2-2p_{22}-s}(\rho_2)}{J_{n_2+r_{2}-2p_{22}}(\rho_2)} + \frac{J_{r_2-s}(\rho_2)}{J_{r_2}(\rho_2)} \Big)  \Big].
\end{multline}
If $k_\perp|x_+^0|\ll1$, then
\begin{equation}
    j_{m- n_1 -n_2 - 2 ( r_1+r_2 -p_{11} -p_{22})}^0\approx\de_{m,n_1 +n_2 + 2 ( r_1+r_2 -p_{11} -p_{22})},
\end{equation}
and we arrive at the selection rule
\begin{equation}\label{sel_rul_2planar}
    m+n_1+n_2 \text{ is an even number}.
\end{equation}
In this case, assuming that the ratio of undulator frequencies, $\eta_2$, is a rational number and introducing the principal quantum number \eqref{Dioph_eqn_sol}, we have
\begin{multline}\label{2omegaAmpl_plane_simpl}
	\mathcal{A} = 2\pi\sum_{n=\sgn(\omega)}^{\sgn(\omega)\infty} \de_N\big(k_0(1-\tilde{n}_3\be_3)- n\omega \big)
    \sum_{k,r_{1,2},p_{11,22}=-\infty}^\infty
    \de_{m, n_1 +n_2 + 2 ( r_1+r_2 -p_{11} -p_{22})} \times \\
	\times  (-1)^{r_2}e^{ik_3z_0}\prod_{i=1}^{2} \Big[ J_{n_i+r_i-2p_{ii}}(\rho_i) J_{r_i}(\rho_i)
    J_{p_{ii}}(\vk_{ii})e^{in_i\chi_i}\Big]\times\\
    \times \Big[ \be_3 -\frac{\omega_1\rho_1}{k_\perp n_\perp}\Big(\frac{J_{n_1+r_1-2p_{11}-s}(\rho_1)}{J_{n_1+r_1-2p_{11}}(\rho_1)} -\frac{J_{r_1-s}(\rho_1)}{J_{r_1}(\rho_1)} \Big) -\frac{\omega_2\rho_2}{k_\perp n_\perp}\Big(\frac{J_{n_2+r_2-2p_{22}-s}(\rho_2)}{J_{n_2+r_2-2p_{22}}(\rho_2)} + \frac{J_{r_2-s}(\rho_2)}{J_{r_2}(\rho_2)} \Big)  \Big],
\end{multline}
where $n_{1,2}$ are expressed in terms of $k$ according to formula \eqref{Dioph_eqn_sol}. For a fixed $n$, the radiation amplitude \eqref{2omegaAmpl_plane_simpl} describes a coherent superposition of states of twisted photons with different $m$ obeying the selection rule \eqref{sel_rul_2planar}, and the phases
\begin{equation}\label{phase_planar}
    n_1\chi_1+n_2\chi_2.
\end{equation}
Unlike a helical two-frequency undulator, in this case $n_1$ and $n_2$ and, consequently, the phase \eqref{phase_planar} are not uniquely defined by the quantum numbers $m$ and $n$.

Let us consider the restrictions that are imposed by the selection rule \eqref{sel_rul_2planar} on the admissible values of TAM projection $m$ at the radiation harmonic with principal quantum number $n$ in the case when $\eta_2$ is a rational number. Substituting \eqref{Dioph_eqn_sol} into \eqref{sel_rul_2planar}, we obtain
\begin{equation}\label{sel_rul_2planar1}
    m+n(n_1^0+n_2^0) +(\la_1-\la_2)k \text{ is an even number}.
\end{equation}
In particular, if the principal quantum number $n$ is even and $(\la_1-\la_2)$ is even, i.e., $\la_1$ and $\la_2$ are odd, then $m$ must be even. This selection rule is generalized to the case of an elliptic $M$-frequency undulator with rational $\eta_i$. Indeed, in this case, the relation \eqref{sel_rule_gen} is satisfied. Substituting expression \eqref{Dioph_eqn_M_sol} for the numbers $n_i$ in \eqref{sel_rule_gen}, we deduce
\begin{equation}\label{sel_rul_planar}
    m+n\sum_{i=1}^M n_i^0 +\sum_{a>b=1}^{M} (\la_a-\la_b) k_{(a,b)} \text{ is an even number}.
\end{equation}
Thus we see that when $n$ is even and all $(\la_a-\la_b)$ are even, the TAM projection of twisted photons, $m$, emitted at a given harmonic must be even. All the differences $(\la_a-\la_b)$ are even for mutually prime $\la_a$ only when all $\la_a$ are odd. The fulfillment of the selection rules \eqref{sel_rul_2planar}, \eqref{sel_rul_planar} can be seen in Fig. \ref{Plot_2planar}.

\section{Conclusion}

The general theory of radiation of twisted photons by $M$-frequency undulators developed in the present paper shows that these undulators can be employed as a bright source of photons prepared in the states that are a linear combination of modes with definite projections of TAM. The amplitude and the phase of coefficients of this linear combination and the frequency of photons are readily managed by changing the parameters of multifrequency undulator. The most promising in this regard are the helical multifrequency undulators, whose magnetic field has the form \eqref{magnetic_field} with $|H^i_x|=|H^i_y|$, as these parameters of photon states are easier to control compared to the elliptical undulators of a general form, in particular, the multifrequency undulators composed of two planar one-frequency undulators, where the magnetic field has the form \eqref{H_field_comp_two_planar}.

The photons in such states were used in the papers \cite{Stock2015,Sherwin2017,Sherwin2020,Ivanov2022} in studying the interference effects in various processes of quantum electrodynamics. As it has been discussed in Introduction, in a scattering process where the initial state contains only one particle with definite projection $m$ of TAM, whereas the other particles in the initial and final states possess the definite momenta, the inclusive transition probability is independent of the phase of this twisted state \cite{Ivanov2022}. If the initial state of the particle participating in the scattering process is a superposition of modes with different values of $m$, then the transition probability will depend on the differences between these values \cite{Stock2015,Sherwin2017,Sherwin2020,Ivanov2022} that gives rise to nontrivial effects missing in scattering of particles prepared in the states with definite projections of momenta.

In Sec. \ref{Probab_Rad_Tw_Phot}, we have derived rather simple general expression \eqref{totAmpl2} for the one-particle amplitude of radiation of a twisted photon by a $M$-frequency undulator. This expression implies, in particular, the selection rule \eqref{sel_rule_gen} for the TAM projection that generalizes the selection rule for radiation of twisted photons by an elliptical one-frequency undulator \cite{KazRyakElUnd}. The reflection symmetry of the radiation probability \eqref{refl_sym_2} has been proved in changing the chirality of the undulator. Besides, we have shown that the radiation probability is invariant under shifting by $\pi$ of all the phases of particle oscillations $\chi_i$. The radiation energy spectrum of $M$-frequency undulator with ratios of the frequencies $\omega_i$ being rational numbers is equidistant with the base frequency in the comoving frame, $|\omega|$, specified by formula \eqref{effective_freq}. This frequency is strictly less than $\min|\omega_i|$ save the case when all $\omega_i$ are multiples of some frequency from the set $\{\omega_i\}$.

As the example, the radiation from two-frequency undulators has been considered. We have investigated the undulator being the composition of $M$ helical one-frequency undulators and the undulator composed of two planar one-frequency undulators with orthogonal planes of oscillations (see, e.g., \cite{Zhukovsky2017,Zhukovsky2017epl,Mirian2014,Datolli2014}). As far the helical $M$-frequency undulator is concerned, the general selection rule \eqref{sel_rul_circ} with respect to TAM projection of photons radiated at a given $n$-th harmonic has been established and the physical interpretation to this selection rule in terms of virtual photons mediating between the charged particle and the undulator has been given. In the particular case of a two-frequency undulator (its magnetic field has the form \eqref{magnetic_field} with $M=2$), simple expression \eqref{2omegaAmpl_simpl1} for the amplitude of twisted photon radiation has been obtained from which it follows that the radiation at a given harmonic is a superposition of modes with definite projections of TAM \eqref{sel_rul2-1}, with relative phases \eqref{rel_phase_circ}, and with amplitudes proportional to the expression on the second line of \eqref{2omegaAmpl_simpl1}. These parameters of the photon state are completely determined by the parameters of the undulator and can be easily controlled. We have found restrictions \eqref{n_i_restrictions} on the possible spectrum of values of $m$ at the $n$-th harmonic implying that the main contributions come from such projections of TAM that correspond to the exchange by order $|n|$ or less of virtual photons with undulator. As a consequence, the spectrum with respect to $m$ at fixed and not very large harmonic number $n$ in a helical two-frequency undulator is comprised of no more than two or three values. One can achieve an increase of the number of realizable values of $m$ at a given $n$ by increasing the number $M$ of frequencies of the undulator (see Fig. \ref{Plot_3helical237}). In the case of a helical two-frequency undulator, where one of the subundulators operates in the dipole regime, the radiation spectra with respect to energy and TAM projection have been described in detail. The scheme given can be readily generalized to the case of a helical $M$-frequency undulator with $M_d<M$ subundulators working in the dipole regime.

As for the composition of two planar one-frequency undulators \eqref{H_field_comp_two_planar}, the explicit expression for the amplitude of radiation of twisted photons and the selection rule for the TAM projections at a given $n$-th harmonic have also been found. The resulting expression for the radiation amplitude is rather huge and the expressions for the relative phases and the amplitudes of coefficients of linear combination of modes with different $m$ and fixed energy are quite complicated. Moreover, we have found the selection rule with respect to $m$ for an arbitrary $M$-frequency undulator at a given $n$-th harmonic. As a result, the condition on the ratios of frequencies in multifrequency undulator has been found that provides only even $m$ for even $n$. In that case, the radiation at such harmonics consists of twisted photons with nonzero projection of the orbital angular momentum and, consequently, the intensity of on-axis radiation for these harmonics vanishes.

As is well-known \cite{BaKaStrbook,Bord.1}, the properties of undulator radiation in the ultrarelativistic regime, $\gamma\gg1$, coincides with the properties of radiation from a charged particle moving in the corresponding laser wave. The virtual photons constituting the field of undulator become very close to the real ones in the rest frame of the charged particle. Therefore, one may consider as a good approximation that the ultrarelativistic charged particle moves in the field of real photons constituting a plane electromagnetic wave. This reasoning implies that the theory developed in the present paper is applicable to description of radiation of twisted photons from charged particles propagating in a plane electromagnetic wave being a superposition of codirectional plane waves with different frequencies and polarizations. A detailed derivation of the amplitudes and the probabilities of radiation of twisted photons in this case will be given elsewhere. As was shown in \cite{KazMokrRyak2025}, the twisted photons can be employed for production of twisted electrons by means of surface photoelectric effect. In this process, the photons prepared in the states that are a superposition of twisted modes appear to generate the electrons in the states with the same spectrum of TAM. Transition radiation from Gaussian beams of charged particles traversing locally isotropic \cite{Takabayashi2025,BKL5} and helical \cite{BKL5,BKKL21,BKKL21pre} media consists of twisted photons. It is clear from the general analysis presented in \cite{Bogdanov2020} that transition and Vavilov-Cherenkov radiations from helically microbunched multifrequency beams with rational ratios of frequencies generate photons in the superposition states that we have investigated in the present paper. We postpone the study of these effects to future research. Notice also that the procedure similar to that has been developed in deriving the amplitude \eqref{totAmpl} can be used to derive the one-particle amplitude of radiation of plane wave photons in $M$-frequency undulator. However, the exposition of this procedure deserves a separate paper.

\paragraph{Acknowledgments.}

The reported study was supported by the Ministry of Science and Higher Education of the Russian Federation, the contract FSWM-2025-0007.


\end{document}